\documentclass[12pt]{article}
\usepackage[a4paper,textwidth=16cm,textheight=22.5cm,footnotesep=\baselineskip]{geometry}
\usepackage{bm}      
\usepackage{amsmath} 
\usepackage{cite}    
\usepackage{slashed}
\usepackage{hyperref}
\hypersetup{final}
\usepackage[utf8]{inputenc}
\usepackage{mathrsfs} 

\IfFileExists{nowidow.sty}{\usepackage[defaultlines=2,all]{nowidow}}{}

\def\gsim{\mbox{~{\raisebox{0.4ex}{$>$}}\hspace{-1.1em}
        {\raisebox{-0.6ex}{$\sim$}}~}}
\def\lsim{\mbox{~{\raisebox{0.4ex}{$<$}}\hspace{-1.1em}
        {\raisebox{-0.6ex}{$\sim$}}~}}

\newcommand{\con}{ _ { _ { \! \rm C } } }
\newcommand{\conleft}{ _ { _ { \! \! \rm C } } }
\newcommand{\LeR}{\ensuremath{L _ { e \rm R }}}
\newcommand{\df}{\ensuremath{\delta\mkern-2mu f}}
\newcommand{\dF}{\ensuremath{\delta\mkern-2mu F}}
\newcommand{\gvec}[1]{\mbox{\boldmath  $#1$}}
\newcommand{\dM}{\ensuremath{\delta\mkern-1mu M}}

\newcommand{\ns}{n _ { \rm s } }
\newcommand{\dX}{\ensuremath{\delta\mkern-1.5mu X}}
\newcommand{\order}{\mathcal O}
\newcommand{\omuv}{\omega _ {\rm fast}}
\renewcommand{\vec}[1]{{\bf #1}}
\newcommand{\T}{\mathcal{T}}
\newcommand{\tpf}{\Delta}

\begin{document}
\setlength{\baselineskip}{0.6cm} 

\makeatletter \@addtoreset{equation}{section} \makeatother
\renewcommand{\theequation}{\arabic{section}.\arabic{equation}}

\newcommand*\xbar[1]{%
  \hbox{%
    \vbox{%
      \hrule height 0.5pt 
      \kern0.5ex
      \hbox{%
        \kern-0.1em
        \ensuremath{#1}%
        \kern-0.1em
      }%
    }%
  }%
} 

\begin{centering}

  \textbf{\Large{Kinetic equations for sterile neutrinos\\[2.5mm]
                 from thermal fluctuations}}

\vspace*{.6cm}

Dietrich B\"odeker%
\footnote{bodeker@physik.uni-bielefeld.de%
} and Dennis Schr\"oder%
\footnote{dennis@physik.uni-bielefeld.de%
}

\vspace*{.6cm} 

{\em 
  Fakult\"at f\"ur Physik,
  Universit\"at Bielefeld,
  33501 Bielefeld,
  Germany
}

\vspace{10mm}

\end{centering}

\vspace{5mm}
\noindent

\begin{abstract}
  We obtain non-linear kinetic equations for sterile neutrino
  occupancies  and  lepton minus baryon  numbers  by matching 
  real time  correlation functions  of  thermal  fluctuations 
  computed in an  effective description  to those computed in 
  thermal quantum field theory.
  After  expanding  in the sterile-neutrino Yukawa couplings, 
  the coefficients in the equations are written as  real time
  correlation functions of Standard Model operators.
  Our kinetic equations are valid for an  arbitrary number of
  sterile neutrinos of any mass spectrum.
  They can be used to describe, e.g.,  low-scale leptogenesis
  via neutrino oscillations, or sterile neutrino  dark matter
  production in the Higgs phase. 
\end{abstract}

\section {Introduction}
\label{s:intro}

Sterile neutrinos may play a key role in the evolution of the
Universe, e.g.~by producing the matter-antimatter
asymmetry~\cite{Fukugita:1986hr} or by
constituting all or part of the dark matter~\cite{Dodelson:1993je}. 
They appear in the arguably 
most straightforward extensions of the Standard Model
which can account for non-zero masses of the active neutrinos. 
Having no Standard Model gauge interactions and typically small Yukawa
couplings, the sterile neutrinos
equilibrate only slowly, if at all, 
so that they can provide the non-equilibrium
conditions required for baryogenesis.

Depending on the masses and couplings of the sterile neutrinos,
leptogenesis can be realized in different stages of the evolution 
of the Universe. 
Thermal (or high-scale) leptogenesis requires very heavy sterile neutrinos,
with masses larger than $ 10 ^ 6 $~GeV~\cite{Moffat:2018wke}.
If there are two nearly mass-degenerate sterile neutrinos,
this bound can be lowered to  $ 10 ^ 3 $~GeV~\cite{Pilaftsis:2003gt}.
While sterile neutrinos with such large masses are 
well motivated by the see-saw mechanism, they are not detectable in
any foreseeable experiment.
Leptogenesis through oscillations~\cite{Akhmedov:1998qx,Asaka:2005pn}
(or low-scale leptogenesis)
can work for even smaller masses, 
below $ \sim 5 $~GeV, and these sterile neutrinos
could in principle be experimentally detected~\cite{Alekhin:2015byh}.

Leptogenesis through oscillations has been described by Boltzmann
equations, and in the relativistic case with several flavors with
generalizations thereof~\cite{Sigl:1992fn}. 
The momentum spectrum of sterile neutrinos is non-thermal 
and it can be important to keep the full momentum 
dependence~\cite{Asaka:2011wq}, but for parameter-space scans
usually momentum space averages are considered (for recent work see,
e.g.,~\cite{Eijima:2018qke,Abada:2018oly}).
There have been various approaches which start from first principles
to avoid some ad-{hoc} assumptions inherent in the Boltzmann equation,
and to systematically include medium 
effects~\cite{Beneke:2010dz,Gagnon:2010kt,Drewes:2012ma,Drewes:2016gmt}.

The approach of~\cite{Bodeker:2014hqa} makes use of 
the slowness of the sterile neutrino's interaction right from the start.
Since most degrees of freedom equilibrate much faster, 
one needs to account for only a few non-equilibrium quantities. 
They can be described by kinetic equations with coefficients
which only depend on the temperature and on the chemical potentials
of conserved charges. 
The coefficients can be expressed in terms of finite temperature
correlation functions~\cite{Bodeker:2014hqa}.
These relations are valid to all orders in Standard Model couplings,
and can thus be used to compute higher order corrections allowing
to estimate the accuracy of the approximations~\cite{Bodeker:2017deo}.
In~\cite{Bodeker:2014hqa} the washout rate  
was obtained by matching thermal fluctuations
in the effective description and in quantum field theory.
It was applied to the  production of a single sterile-neutrino 
species in~\cite{Bodeker:2015exa}. 
Non-linear terms in the kinetic equations
may also play an important role~\cite{Asaka:2011wq}. 
In~\cite{Ghiglieri:2017gjz} the rates 
for leptogenesis including non-linear terms were
obtained using a quite different approach. There the sterile-neutrino
masses were neglected. 
In~\cite{Ghiglieri:2019kbw} non-linear equations were obtained also for
the massive case. 

In this paper we generalize the approach of~\cite{Bodeker:2014hqa} to
several species of sterile neutrinos which allows 
for oscillations, and by including non-linear terms.
It is organized as follows.  In section~\ref{s:kinetic}
we obtain master formulae for the coefficients in the equations of motion
in terms of correlators in thermal field theory.
In section~\ref{s:N} we define
the Lagrangian and the variables used
to describe the sterile neutrino densities, and we obtain kinetic
equations in terms of correlators of Standard Model operators.
In section~\ref{s:charges} we discuss the conserved charges
in the various temperature regimes
and their connection with the chemical potentials in the evolution equations,
elaborating on the role played by the lepton number carried by
right-handed electrons.
We give the expressions for the dissipative contribution from multiple soft scattering
and compute some of the dispersive contributions, both 
in the symmetric phase, in section~\ref{s:sm}, and
we summarize our results in section~\ref{s:sum}.
We describe the perturbative solution of the equations
of motion for fluctuations
in appendix~\ref{a:pert}, illustrate the coarse-graining
method in momentum space needed to apply our method to occupancies
in appendix~\ref{a:smear}, summarize some useful
formulae concerning Green's functions in appendix~\ref{a:ana}, and
demonstrate the cancellation of some rate coefficients
in appendix~\ref{a:gfff}.

\paragraph{Notation} We write four-vectors in lower-case italics, $ k $,
and the corresponding three-vectors in boldface, $ \vec k $. Integrals
over three-momentum are denoted by $ \int _ { \vec k } \equiv
 ( 2 \pi ) ^ { -3  } \! \int \! d ^ 3 k $. 
Four-vectors are denoted as  $ k = ( k ^ 0 , \vec k ) $. 
In imaginary time the Matsubara frequencies are 
$ k ^ 0 = i  2 n  \pi  T $
for bosonic or $ k ^ 0 = i (2 n + 1  ) \pi  T $ for fermionic operators
with integer $ n $, $ T $ is the temperature.
We denote fermionic Matsubara sums by a tilde,~%
$ \widetilde { \sum } _ { k ^ 0 } $.
We use the metric with signature $ (+,-,-,-) $.

\section{Kinetic equations and real time correlations}
\label{s:kinetic}

We consider an out-of-equilibrium system in which some quantities 
evolve much more slowly than most other degrees
of freedom. Their deviations from equilibrium, which we denote by
$ y _ a $, will depend on time. The non-equilibrium state
is determined by the values of the $ y _ a $, and
by the temperature of the system, as well as the values
of the conserved charges.
Therefore, the time derivative of $ y _ a $ depends
only on these quantities.
We assume that we can expand
\begin{align}
  \dot y _ a 
  = { }  
  - \gamma  _ { ab } \, y _ b
  - \frac 12 \gamma  _ { abc } \, y _ b \, y _ c 
  - \frac 1 { 3 ! } \gamma  _ { abcd } \, y _ b \, y _ c \, y _ d
  - \cdots  
  \label{eom}
  .
\end{align} 
The $ y _ a $ and the rate coefficients $ \gamma  $ are assumed to be real.
The $ \gamma  $'s only depend on the temperature and the values
of the (strictly) conserved charges.
The effective kinetic equations~(\ref{eom})
are valid for frequencies $ {\omega \ll \omuv}$
where $\omuv$ is the characteristic frequency of the `fast,'
or  `spectator'
processes, which keep the other degrees of freedom in
thermal equilibrium.

For sufficiently small $ y _ a $
the equations of motion can be linearized.
This approximation
has been used to obtain equations for 
thermal leptogenesis in~\cite{Bodeker:2013qaa}. 
Non-linear terms in~(\ref{eom}) can be important
even when all $ y _ a $ are small, but hierarchical such that,
e.g., $ y _ 1 $ is of similar size as $ y _ 2 y _ 3 $. 
Non-linear terms in the kinetic equations for
leptogenesis through oscillations were taken into account
in~\cite{Asaka:2011wq}. 
In resonant dark matter production~\cite{Shi:1998km} they naturally occur in
the resummed active neutrino propagator. 

\subsection{Correlators in the effective theory}
\label{s:langevin}

The thermal fluctuations of the slow variables $ y _ a $ satisfy 
the same type of equations as~(\ref{eom}), but with
an additional Gaussian noise term on the right-hand side, 
representing the effect of rapidly fluctuating quantities.%
\footnote{See, e.g., \S 118
   of reference~\cite{Landau:1980mil}
   on correlations of fluctuations in time.}  

From these equations
one can compute real-time correlation functions of the fluctuations such as
\begin{equation}
   \mathcal{C}_{  a    b  }(t)=\langle y_{a  }(t)y_{b  }(0)\rangle
   \label{ccor} 
\end{equation}
by
solving these equations and then averaging over  the noise and over 
initial conditions.%
\footnote{Non-linear terms in the equation of motion could potentially
lead to non-vanishing expectation values of $ y _ a $. 
Therefore in general one also has to include  $ y $-independent
terms on the right-hand side to ensure that the expectation values vanish.
Eventually we want to describe deviations from thermal equilibrium
   with~(\ref{eom}). Then the $ y _ a $ are much larger than their thermal
   fluctuations, and the $ y $-independent term will be small 
   compared to the non-linear terms in~(\ref{eom}) and can be neglected.} 

We solve the equations of motion for the fluctuations
by one-sided Fourier transformation
\begin{align}
  y ^ + _ a  ( \omega  )  
    \equiv 
    \int_{0}^{\infty}dt
  \, e^{i\omega t}
  y _ a ( t )
  \label{laplace} 
  .
\end{align}
At linear order we obtain
\begin{align}
  y _ a ^ { + ( 0 ) }  ( \omega  ) 
  =
  \left[ ( - i \omega  + \gamma  ) ^{ -1 } \right] _ { ab }  
  y _ b
  ( 0 ) 
   + \cdots  
  \label{y0} 
  ,
\end{align}    
where the ellipsis represents a term linear in the noise. 
Inserting this into the one-sided Fourier transform of~(\ref{ccor})
one obtains~\cite{Bodeker:2014hqa}%
\begin{equation}
  \mathcal{C} ^ + _{ a  b }
  ( \omega  ) 
  =
  \left[ ( - i \omega  + \gamma  ) ^{ -1 } \right] _ { ac }   
  \Xi _ { c b }  
  \label{ceff} 
  .
\end{equation}
Here the noise term has dropped out.
When averaging over the
initial conditions at $ t = 0 $, 
one encounters the real and symmetric susceptibility matrix with elements
\begin{align} 
  \Xi_{ab} \equiv \langle y_{a} 
   y_{b}
  \rangle
  \label{Xi} 
   ,
\end{align} 
i.e., the {\it equal }time correlators $ \mathcal{C}_{  a    b  }( 0 ) $.
As in~\eqref{ccor}, the average in~(\ref{Xi}) is
canonical, that is, at fixed
values of the conserved charges.  

The rate matrix $ \gamma _ { a b } $ can be extracted from~\eqref{ceff} by 
considering frequencies $ \omega  $ which are parametrically
much larger than the elements of the matrix $ \gamma $. Then one can 
expand~(\ref{ceff}) in $ \gamma /\omega  $. For real
$ \omega  $ the leading term in this expansion
is purely imaginary. Thus by taking the
real part of~(\ref{ceff}) one can extract the next term
which is linear in $ \gamma  $~\cite{Bodeker:2014hqa},%
\begin{equation}
    {\rm Re } \; 
   \mathcal{C}_{ab}^{+}(\omega)        = 
   \frac 1 { \omega^{2}} \gamma_{ac}
 \, \Xi _{cb}
   + \order ( \omega  ^ { -3 } ) 
    \qquad (\mbox{for real } \omega  ) 
       .
   \label{classic}
\end{equation}
Here it is important that we take the one-sided Fourier
transform instead of the Fourier transform because the latter  only
depends on the symmetric part of $ \gamma $.

Now we go beyond the linear order.
We include the non-linear terms in the equation motion, and expand
\begin{align} 
  y = y ^{ ( 0 ) } + y ^{ ( 1 ) } + y ^{ ( 2 ) } + \cdots  
  \label{ypert}
\end{align}
where $ y ^{ ( n ) } $ is of order 
$ \left (  y ( 0 )  \right ) ^ { n + 1 }$
and vanishes at the initial time $ t = 0 $.
We will encounter the generalization of~\eqref{Xi},
\begin{align}
\Xi _ { a _ 1 a _ 2 \cdots a _ n }
\equiv
\langle y _ { a _ 1 } y _ { a _ 2 } \cdots y _ { a _ n } \rangle
\con
\label{Xihigh}
,
\end{align}
where the subscript `\textrm{C}' indicates that we  only include
the connected part, for which we assume 
\begin{align}
\left( \Xi _ { a _ 1 \cdots a _ m } \right) ^ { 1 / m } 
\ll 
\left( \Xi _ { a _ 1 \cdots a _ n } \right) ^ { 1 / n }
\qquad (m > n \geq 2)
\label{Xihier}
   .
\end{align}
This can be seen as a consequence of our assumption that we
can expand the right-hand side of equation~(\ref{eom}),
since the time evolution of the fluctuations
also determines their equal time correlations
(see, e.g.,~\cite{ZinnJustin:1989mi}).
For the occupancies of sterile neutrinos
we have checked the assumption (\ref{Xihier}) in appendix~\ref{a:smear}.
The coefficient $\gamma  _ { abc } $ of the quadratic term
in equation~(\ref{eom}) can then be extracted from the correlation function
\begin{align}
  \mathcal{C}_{a ( b c ) }(t) 
  \equiv 
  \langle y _ a ( t ) 
  y _ b y _ c 
  ( 0 ) \rangle
  \label{cabc} 
\end{align} 
as follows (for details see appendix~\ref{a:pert}).%
\footnote{Note that~\eqref{cabc} is
connected, because the expectation value of a single $ y $ vanishes.}
We have
\begin{align}
    \mathcal{C}_{a ( b c ) }(t) 
    =
    \left \langle y _ a ^{ ( 0 ) } ( t )
    y _ b y _ c
    ( 0 )
       \right \rangle 
    +
    \left \langle y _ a ^{ ( 1 ) } ( t ) 
    y _ b  y  _ c
    ( 0 ) 
    \right \rangle
    \label{3pt}
    .
\end{align}
The first term on the right-hand side is very similar to~(\ref{ceff}), one only
has to replace~(\ref{Xi}) with the expectation value of three factors
of $ y ( 0 ) $.
Again we take the one-sided Fourier transform.
Our assumption~\eqref{Xihier} allows us to neglect the contribution
from $ \Xi  _ { abcd } $,
which gives
\begin{align} 
  \langle y _ a ^{ + ( 1 ) } ( \omega   ) 
  y _ b y _ c
  ( 0 ) 
    \rangle  \con
    = 
    \frac { 1 } { \omega  ^ 2 } \gamma  _ { a de } \, \Xi  _ { db } \,
    \Xi  _ { ec }
    + \order ( \omega  ^{ -3 } ) 
    .
    \label{con} 
\end{align} 
Thus we obtain
\begin{align} 
    {\rm Re } \; \mathcal{C}_{a ( bc ) }^{+}(\omega)
  =
    \frac 1 { \omega    ^ 2 } 
  \left [ 
  \gamma  _ { ade}  \,
   \Xi_{db} \, \Xi_{ec}
   + \gamma  _ { ai } \, \Xi  _ { i bc } 
 \right ]
  + \order ( \omega ^{ -3 }  ) 
  \label{g3}
  ,
\end{align}
which allows us to extract $ \gamma  _ { abc } $. 
Similarly we obtain the coefficient 
multiplying the cubic term in~(\ref{eom})
by solving the equation of 
motion for $ y _ a $ perturbatively up 
to linear order in $ \gamma _ { abc } $ and $ \gamma _ { abcd } $ 
and successively 
computing the {\it connected} correlation function
\begin{align}
  \mathcal{C}_{a ( b c d ) }(t) 
  \equiv 
  \langle y _ a ( t ) 
  y _ b   y _ c   y _ d 
  ( 0 ) \rangle
  \con
  \label{cabcd} 
  \;
  .
\end{align} 
Following the same line of arguments, we obtain
\begin{align}
    {\rm Re } \; \mathcal{C}_{a ( bcd ) }^{+}(\omega)
  =
    \frac 1 { \omega    ^ 2 } 
  \left [ 
  \gamma  _ { aijk}  \,
   \Xi_{ib} \, \Xi_{jc} \,
\Xi_{kd}
   +   \frac 12 \,  \gamma  _ { aij } \, \Xi  _ { ij bcd } + \gamma  _ { ai } \, \Xi  _ { i bcd } 
 \right ]
  + \order ( \omega ^{ -3 }  ) 
  \label{g4}
  .
\end{align}

\subsection{Correlators in the microscopic theory}
\label{s:mic} 

The one-sided Fourier transforms of the correlation
functions~(\ref{ccor}),~(\ref{cabc}), and~\eqref{cabcd} as well as the 
susceptibilities~(\ref{Xihigh}) can also be computed in the microscopic
quantum theory. In the range of validity $ \omega \ll \omuv $
of the effective equations of motion~(\ref{eom}) they have to
match their counterparts in the effective theory.  
This way the
coefficients in~(\ref{eom}) can be computed
from~(\ref{classic}),~(\ref{g3}) and~(\ref{g4}) with the quantum
correlators on the left-hand side, evaluated in the regime 
$ \gamma\ll\omega\ll\omuv $. 
In this regime $ \mathcal{C}_{ab}^{+} $ has to match 
the one-sided Fourier transform of the microscopic correlation function
\begin{equation}
   C_{ab}(t)
   \equiv 
   \frac{1}{2}\Big\langle \Big\{ y_{a}(t),y_{b}(0)\Big\} \Big \rangle 
   \label{quantum} 
   .
\end{equation}
Since $ \omuv \lsim T $, we are dealing with 
frequencies $ \omega $ much smaller than the temperature.
In this regime the one-sided Fourier transform of~(\ref{quantum}) 
is approximately given by~\cite{Bodeker:2017deo} 
\begin{align}
  C ^ + _ { ab } ( \omega  ) 
  =
  -i \frac T \omega  \big [ \tpf    _ { ab } ( \omega  ) - \tpf    _ { ab }
  ( 0 ) \big ]
  ,
  \label{quantum+} 
\end{align} 
where  
\begin{align} 
 \tpf_{ab}  (\omega  )
 \equiv 
 \int
 \frac{d\omega '}{2\pi} \frac{\rho_{ab}(\omega')}{\omega'-\omega }
 \label{spec}
 . 
\end{align} 
(\ref{spec}) is an analytic function off the real axis, 
and 
\begin{align}
  \rho_{ab}(\omega) 
  & \equiv 
  \int \! dt \, e^{i\omega t}
  \Big\langle \Big[y_{a}(t),y_{b}(0)\Big]\Big\rangle 
  \label{rho}
\end{align}
is the  spectral function of the bosonic operators $ y _ a $ and $ y _ b $.
For real $ \omega $,  
$ \tpf _ { ab } ( \omega + i 0 ^ + ) $
equals the retarded two-point function 
$\tpf ^ { \rm ret } _ { a b } ( \omega )$ (see~(\ref{retarded})).
Matching $ { \cal C } ^ + $ and $ C ^ + $,
and using~(\ref{classic}) as well as the fact that $ \tpf _ { ab } ( 0
) $ is real one obtains the master formula~\cite{Bodeker:2017deo}
\begin{equation}
  \gamma_{ab}
  =
  T \omega \, \text{Im} \,\tpf  _{ac}^ {\rm     ret}(\omega)
 (\Xi^{-1})_{cb}
 \qquad (  \gamma\ll\omega\ll\omuv ) 
 \label{kubo} 
 .
\end{equation}
For real spectral functions it agrees with the 
Kubo-type relation in~\cite{Bodeker:2014hqa}. 
Following the same steps with~(\ref{g3}) and~(\ref{g4}) we obtain
the master formulas
\begin{align} 
  \gamma  _ { abc } 
  &=
  \Big [ T \omega  
  \, {\rm Im \,}  \tpf ^ {\rm     ret}_{a ( de ) }(\omega)
    -  \gamma _ { a f }  \,
    \Xi  _ { fde }
    \Big ]
    (\Xi^{-1})_{db} (\Xi^{-1})_{ec}
  \label{gamma2}
  ,\\[0.2cm]
  \gamma  _ { abcd } 
  &=
  \bigg [  T \omega  
  \, {\rm Im \,}  \tpf ^ {\rm     ret}_{a ( efg ) }(\omega)
    - \frac 12 \gamma  _ { a i j } \, \Xi  _ { i j e f g }
    -  \gamma _ { a i }   \, \Xi  _ { i e f g }
 \bigg ]
 (\Xi^{-1})_{ e b } (\Xi^{-1})_{ f c } (\Xi^{-1})_{ g d } 
  \label{gamma3}
  ,
\end{align}
where in both cases $ \gamma \ll \omega \ll \omuv $.
As in~\eqref{cabcd}, we include only the connected
piece of the correlator $\tpf ^ {\rm     ret}_{a ( efg ) }$.
In general the operators inside the retarded correlators
will not necessarily commute at equal times.

In some cases it is more convenient to compute the correlators of time
derivatives of one or both of the operators, and then use
\begin{align} 
  \tpf _ { AB } ^{ {\rm ret } } ( \omega  ) 
  &=
  \frac 1 { \omega  } 
  \Big [ 
    i \tpf _ { \dot A B } ^{ {\rm ret } } ( \omega  ) 
    + \big\langle [ A ( 0 ) , B ( 0 ) ] \big \rangle 
    \Big ] 
    \label{punkt1} 
    ,
    \\
  \tpf _ { AB } ^{ {\rm ret } } ( \omega  ) 
  &=
  \frac 1 { \omega  ^ 2 } 
  \Big [ 
    \tpf _ { \dot A \dot B } ^{ {\rm ret } } ( \omega  ) 
    + i  \big \langle [ A ( 0 ) , \dot B ( 0 ) ] \big \rangle 
    + \omega \big   \langle [ A ( 0 ) , B ( 0 ) ] \big \rangle 
    \Big ] 
    \label{punkt2} 
    .
\end{align} 

\section{Kinetic equations for sterile neutrinos}
\label{s:N}

We now consider the Standard Model extended by $ \ns $ flavors
of sterile (or right-handed) neutrinos $ N _ i $.
The full Lagrangian of the system is given by
\begin{align}
  \mathscr { L } = \mathscr { L } _ { \rm SM } + 
  \frac 12  
  \xbar { N }
  (
    i \slashed \partial - M
    )
  N
  - 
  (
   \xbar  N
   \, h  
   \, J
    + \mbox{H.c.}
  )
  \label{L} 
\end{align} 
with 
\begin{align} 
   \label{J}  
  J \equiv {\widetilde \varphi}^\dagger \, \ell
  ,
\end{align}
where $\widetilde \varphi \equiv i \sigma ^ 2 \varphi ^ \ast $
with the Pauli matrix $ \sigma ^ 2 $.
We describe the sterile neutrinos by the Majorana spinors $ N _ i $, 
in a flavor basis with diagonal mass matrix $ M $.
In general, the matrix of 
Yukawa couplings $h$ is then non-diagonal with 
elements $ h _ { i \alpha } $ for~$ \alpha = e, \mu , \tau $.
For $ \ns > 1 $ the Yukawa couplings can 
violate CP, with an amount which  may be much larger
than the one in the quark sector of the Standard Model. 
The latter is way too small for 
generating the observed baryon asymmetry of the
Universe\footnote{The quantity $ \Omega _ B h ^ 2 $ 
has been measured by Planck~\cite{Aghanim:2018eyx} to a high precision.
It is related to the quantity in~\eqref{nB}, where $ n_ B $ and $ s $ 
denote baryon number density and entropy density, respectively, via
$ n _ B / s = 3.887 \cdot 10 ^ { - 9 } \, \Omega _ B h ^ 2 $, see, e.g.,
chapter~5.2 of reference~\cite{Gorbunov:2011zz}.}
\begin{align}
\frac { n _ { B } } { s }
=
8.71 (4) \cdot 10 ^ { - 11 }
\label{nB}
,
\end{align}
and we neglect it in the following.

We consider temperatures at which the muon Yukawa interaction is in equilibrium,
which is the case when $ T \ll 10 ^ 9$~GeV~\cite{Barbieri:1999ma}.
Then there are two types of slow variables we are
interested in. The first type are the charges 
\begin{align} 
   X _ \alpha  \equiv L _ \alpha  - B /3
   \label{Xalpha}
\end{align} 
where $ L _ \alpha  $ is the lepton number in flavor $ \alpha  $
and $ B $ is the baryon number. 
Unlike $ L _ \alpha  $ and $ B $, the conservation of~(\ref{Xalpha}) 
is not  violated by the chiral anomaly, so that sphaleron processes
do not change~(\ref{Xalpha}).
In the presence of conserved charges the $ X _ \alpha  $
can have a non-vanishing equilibrium value $ X _ \alpha  ^ { \rm eq } $,
so that the $ y _ a $ in~\eqref{eom} correspond to 
$ \dX _ \alpha \equiv X _ \alpha - X _ \alpha ^ { \rm eq } $.
We discuss the equilibrium expectation values in section~\ref{s:charges}.
At $ T \sim 85 $~TeV,
when the rate of  electron Yukawa interactions is 
comparable to the Hubble rate~\cite{Bodeker:2019ajh}, 
the lepton number carried by right-handed electrons $ \LeR $ is a slow 
variable, 
and at $ T \sim 130 $~GeV, when electroweak sphalerons 
freeze out~\cite{DOnofrio:2014rug},
baryon number is slow as well.
The conservation of 
$ L _ { e \rm R } $ and $ B $
is not violated by the sterile-neutrino 
Yukawa interaction. 
However, the $ X _ { \alpha } $, $ L _ { e \rm R } $ and $ B $ are 
individually correlated with U(1)-hypercharge,
so that their evolution equations do not decouple.
When $ T \gsim 10^9$~GeV
the muon Yukawa coupling causes slow interactions and
additional, flavor-non-diagonal charges have to be 
taken into account~\cite{Beneke:2010dz}.

We consider a finite
volume $ V $ and take   $ V \rightarrow \infty $ in the end.
Without the Yukawa interaction, $ N $ would be a free field and 
the equation of motion would give
\begin{equation}
     N_i(x) 
   = 
    \sum_ { \vec k \, \lambda  } 
    \frac 1 { \sqrt{ 
        2  E _ { \vec k  i}  V
      }
    }
   \left [e ^{ i \vec k \vec x  } \,   u_{ \vec k  i \lambda }   \, 
     a _ { \vec k i \lambda } ( t ) 
     +
      e ^{- i \vec k \vec x } \, v_{ \vec k i \lambda}  \, 
      a ^\dagger _ { \vec k i \lambda  } ( t ) 
     \right ]
     \label{NI}
\end{equation}
with $ a _ { \vec k i \lambda} ( t ) = \exp ( - i E _ { \vec k i } t ) a _ {
  \vec k i \lambda} ( 0 ) $ and $ E _ { \vec k i} =
  (  { \vec k  } ^ 2 + M _ i ^ 2 ) ^ { 1 / 2 } $.
The spinors $ u $ and $ v $ are chosen such that 
$  a ^ \dagger _ { \vec k  i \pm } $ creates a sterile
neutrino with helicity $ \pm 1/2 $.
The sterile neutrinos can not be expected to be in kinetic equilibrium
since kinetic and chemical equilibration
are due to the same processes.  
Therefore the other type of slow variables consists of the 
phase space densities, or occupancies, of the sterile neutrinos.
For each $ \vec k $ and $ \lambda  $ the occupation number operators 
form a matrix, called matrix of densities,
or density matrix,  with elements\footnote{In the literature 
  there are several conventions for the order of the indices. We use the one
  of~\cite{Ghiglieri:2017gjz}.}
\begin{align}
  ( f _ { \vec k \lambda   } ) _ { ij } 
  & \equiv
  a ^\dagger   _ {  \vec k i \lambda   } 
  a   _ {  \vec k  j \lambda  }
  \label{fij}
  .
\end{align}
In the presence of the Yukawa interaction in~(\ref{L}) we {\it define} 
the occupation number operators through equations~(\ref{NI}) and~(\ref{fij}).%
\footnote{%
This definition slightly differs from the one in~\cite{Bodeker:2015exa}.
The definition in~\cite{Bodeker:2015exa} and our present definition are
equivalent to the first and the second definition
in~\cite{Ghiglieri:2017gjz}, respectively. 
} 
Their equilibrium values read
\begin{align}
  ( f _ { \vec k \lambda  } ^ { \rm eq } ) _ { i j } ^ { \vphantom {eq}}
  &=
  \delta _ { i j } f _ { \rm F } ( E _ { \vec k i }  ) 
  \label{fijeq}
  ,
\end{align}
with the Fermi-Dirac distribution
$ f _ { \rm F } (E) \equiv 1/(e^{E/T}+1)$. 

The variables appearing in the effective kinetic equations~(\ref{eom}) are real.
Therefore we consider the Hermitian operators 
\begin{align} 
  f ^ a _ {  \vec k \lambda   }
  & \equiv
  T ^ a _ { ij } 
  a ^\dagger   _ {  \vec k i \lambda   } 
  a   _ {  \vec k  j \lambda  }
  \label{fa} 
  .
\end{align}
The  $ T ^ a $ are the  Hermitian U($ \ns $) generators 
satisfying the normalization and completeness
relations
\begin{align}
{\rm tr} (  T ^ a T ^ b ) = \frac { \delta ^ { a b } }  { 2 },
\qquad\qquad
{\textstyle \sum} _ a
T ^ a _ { i j } T ^ a _ { k l } = \frac { \delta _{ i l } \delta _{ j k }} {2}
\label{Tnorm}
.
\end{align}
We write $ \df \equiv f - f ^ { \rm eq } $ for both  $ (f_ {\vec k \lambda } ) _ {ij}$
and $ f ^ a _ { \vec k \lambda } $. The $ \df ^ a _ { \vec k \lambda } $ 
appear as slow variables $ y _ a $ in the equations of
motion~\eqref{eom}.

We will expand the kinetic equations~\eqref{eom} to order $ h ^ 2 $,
and to second order in the deviations $ \dX $
of the charges~(\ref{Xalpha}), 
including the terms of order $ ( \dX ) ^ 2 \df $.
Terms with more than one factor of $ \df $ do not enter
the kinetic equations at order $ h ^ 2 $, as we show in appendix~\ref{a:gfff}.

The fluctuations of the occupancies are comparable to the deviation of
$ f $ from equilibrium, and the higher susceptibilities~\eqref{Xihigh}
of  $ f $ do not satisfy~\eqref{Xihier}.
Strictly speaking, the theory in the preceding
sections is therefore not applicable to the occupancy.
However, one can coarse-grain the operators $ f $ over a certain
momentum region, and the resulting operators satisfy the requirements
of the framework developed in the above sections. The dependence on the
momentum averaging volume drops out in the end. This way we
can effectively use the original operators $ f $ instead of their
smeared versions in our equations.
We elaborate on the details of this procedure in appendix~\ref{a:smear}.

\subsection{Correlation functions}
\label{s:hexp}

With a certain degeneracy of the vacuum masses in~(\ref{L}),
the sterile neutrinos undergo oscillations, which appear already at
order $ h ^ 0 $.
They are described by the off-diagonal matrix
elements in~(\ref{fij}).
One can obtain the equation of motion simply by taking the time derivative
of the operators~(\ref{fij}) and taking the expectation value. 
However, it is also instructive to use the 
Kubo relation~(\ref{kubo}). 
The equilibrium contribution cancels the
disconnected contractions, such that we can replace  $ \df \to f $ 
in~\eqref{ff2pt} and consider only the connected two-point function
\begin{align}
\tpf _ { f ^ a _ { \vec k \lambda } f ^ b _ { \vec p \lambda ' } }
(t)
\equiv
  \big \langle \T  f ^ a _ {  \vec k \lambda  }  ( t  ) 
                   f ^ b _ { \vec p \lambda ' }  ( 0 ) 
   \big \rangle 
   \con
\label{ff}
.
\end{align}
The time $ t $ in~(\ref{ff}) is imaginary, $ t = -i \tau  $ 
with  real $ \tau $, and
$ \T $ denotes time ordering with respect to $ \tau  $. 
We encounter the 2-point functions of the operators appearing
in~(\ref{NI}), for which we find (for both positive and negative $ \tau $)
\begin{align}
\big\langle
\T
a ^ \dagger _ { i \vec k \lambda  } ( - i \tau )
a _ { j \vec p \lambda' } ( 0 )
\big\rangle
=
\delta _ { i j } \,
\delta _ { \vec k \vec p } \,
\delta _ { \lambda \lambda' } \,
T \widetilde {\sum \limits _ { p ^ 0 }}
\frac { e ^ { p ^ 0 \tau } } { p ^ 0 - E _ { i \vec k } }
\label{aprop}
.
\end{align}
The retarded correlators appearing
in~(\ref{kubo}) are obtained
by Fourier transforming~(\ref{ff})   with imaginary
bosonic Matsubara frequency $ \omega  = i \omega _ n \equiv  i 2 \pi  n T$, and
then continuing $ \omega   $ to
the real axis,
$ { \tpf ^ { \rm ret } ( \omega ) = \tpf ( \omega + i 0^+ ) } $
with real~$ \omega $.
One encounters factors like 
$ 1/ ( \omega  + E_ { \vec k i } - E _ { \vec k j } ) $.
To give a contribution to~(\ref{kubo})
this has to be approximately $ 1/\omega  $ when $ \omega  \ll \omuv  $.
This requires
\begin{align}
| \dM_ { i j } ^ 2 | /E _ { \vec k i } \sim 
| \dM _ { ij } ^ 2 | /T 
\ll \omega
\label{degen}
\end{align}
with
\begin{align}
\dM ^ 2 _ { i j } \equiv M ^ 2 _ i - M ^ 2 _ j
\label{deltaM2}
,
\end{align}
which means that the frequency for the oscillations between
sterile flavors $ i $ and $ j $ has to be small compared to~$ \omuv  $.
After a simple computation we obtain at order~$ h ^ 0 $
\begin{align}
  \omega\,  
   \tpf _ { f ^ a _ { \vec{k} \lambda } f ^ b _ { \vec{p} \lambda ' } } 
     ^ { \rm ret } ( \omega ) 
  = { }
  - 
  \delta  _ { \vec k \vec p } \, \delta _ { \lambda \lambda ' }
  \sum _ { ( ij ) }  
  f ' _ { \text F } ( E _ {\vec k i } )
 \frac{ \dM ^ 2 _ { ij }} { 2 E _ {\vec k i}  }   
\,
   \,
   T ^ a _ { ij } T ^ b _ { ji } 
  +
  \order \big( \omega, h^2, \dM ^ 4 \big)
  \label{retffM}
  ,
\end{align}
The notation $ ( i j ) $ indicates that we only sum
over indices with 
$ | \dM _ { i j } ^ 2 | / T \ll \omuv $.
After expanding in $ \dM ^ 2 $
and $ h $, the retarded correlator
entering~\eqref{kubo}
no longer knows about the scale $ \gamma  $, and we can put  $ \omega \to 0 $.

At order $ h ^ 2 $ only the 
first terms in square brackets in~\eqref{gamma2}
and~\eqref{gamma3} survive in the kinetic
equations~\eqref{fkineq} and~\eqref{Xkineq}, if we expand to quadratic
order in chemical potentials. Therefore
we only discuss these terms in the following.
Terms with $ \Xi _ { XXX } $ are canceled once the slow charges
are expressed through their chemical potentials, see~\eqref{muXiX},
and the terms containing $ \Xi _ { f f f } $ or $ \Xi _ { ffff } $
lead to cancellation of the coefficients
$ \gamma  _ { fff } $ and $ \gamma  _ { ffff } $,
which we demonstrate in appendix~\ref{a:gfff}.
Since they are defined as the \emph{connected} pieces, the contributions from the
susceptibilities~\eqref{Xihigh} with mixed indices $ f $ and $ X $ vanish
at leading order in~$ h $, since the correlators they multiply in the master formulae
are already of order~$ h ^ 2 $.

We keep the dependence on the absolute Majorana mass scale in all expressions.
This is important
in order to be able to obtain kinetic equations describing light sterile neutrino
dark matter production during the QCD epoch, because here the $\order ( h ^ 2 )$
terms actually go like $ h ^ 2 M ^ 2 $~\cite{Ghiglieri:2015jua}. 
This dependence emerges from
the retarded active neutrino self-energy, which appears in the
kinetic equations in the Higgs phase.
At order $ h ^ 2 $ we neglect terms of order $ \dM ^ 2 $,  because
both $ h ^2 $ and $ \dM^2$ are small quantities.

To determine the coefficients $ \gamma _ { ff } $, $ \gamma _ { ff X } $,
and $ \gamma _ { ff XX } $ we employ~(\ref{kubo}) through~(\ref{gamma3})
directly, without making use of~\eqref{punkt2}. The latter relation
turns out to be very inconvenient here if there is more than one 
sterile flavor due to
an UV divergent contribution  from the
commutators, which only cancels against a divergence in the
first term in square brackets of~(\ref{punkt2}).
Thus we consider
\begin{align}
  \tpf _
    { f ^ a _ {  \vec k \lambda  } 
     ( f ^ b _ {  \vec p \lambda ' }   X _ { \alpha _ 1 }
                                   \cdots  X _ {\alpha  _ n } ) } 
  ( t ) 
  \equiv 
  \big \langle \T  f ^ a _ {  \vec k \lambda  }  ( t  ) 
  \left ( f ^ b _ { \vec p \lambda ' } 
   X _ { \alpha _ 1 } \cdots  X _ {\alpha  _ n }  \right ) ( 0 ) 
   \big \rangle 
   \con
  \label{ff2pt}
  .
\end{align}
Again, we were able to replace $ \df \to f $ and here also $ \dX \to X $
on the right-hand side of~\eqref{ff2pt}, since we need only the connected
correlator.
We adopt this procedure in the remainder of
this section, understanding that all expectation values are connected.
As in~\eqref{ff}, $ t $ is imaginary.
$  f ^ b _{  \vec p \lambda ' }  $ and the charge operators
$ X _ {\alpha _ i } $ commute at equal times.
The $ \order ( h ^ 2 )$ contribution to   \eqref{ff2pt} becomes
\begin{align}
  \tpf _
    { f ^ a _ {   \vec k \lambda  } 
    ( f ^ b _ {   \vec p \lambda  }   X _ {\alpha _ 1} 
                             \cdots X _ { \alpha _ n } ) } 
&
  (t ) 
    =
    \int \! d ^ 4 x _ 1 d ^ 4 x _ 2 \;     
   \label{fac} \\
   \times \,
   {\rm tr }
\Big \{ h ^\dagger
   \Big \langle \T
   &
   N ( x _ 2 ) \xbar N ( x _ 1 ) f ^ a  _{  \vec k \lambda  }  ( t )
   f ^ b _ {  \vec p \lambda  }  ( 0 )
  \Big\rangle \conleft  \,
    h  \; \Big \langle \T J ( x _ 1 ) \bar J ( x _ 2 ) 
   ( X _{ \alpha _ 1} \cdots  
    X _ {\alpha _ n } ) ( 0 ) \Big  \rangle \conleft
   \; \Big \}  
   \nonumber
   ,
\end{align} 
where the trace refers to both spinor and active flavor indices.
Since we consider the leading order in $ h $,
we can neglect the sterile-neutrino Yukawa interaction in the expectation values
on the right-hand side of~(\ref{fac}).
Our definition of $ a $ and $ a  ^\dagger $
allows us to substitute them for $ N $ and $ \xbar N $
in the path integral, and then work with~\eqref{aprop}.
The second expectation value in~\eqref{fac} is now a correlation function
containing only Standard Model fields.
When $ h $ is neglected,
the charges $ X _ \alpha  $ are conserved, and one 
can introduce
chemical potentials $ \mu  _ { X _ \alpha  } $ such that
\begin{align} 
  \Big \langle \T J ( x _ 1 ) \bar J ( x _ 2 ) 
  ( X _ { \alpha  _ 1} 
  \cdots
  X _ { \alpha  _ n} ) ( 0 ) \Big  \rangle  \conleft
  =
  \left [ 
  T \frac \partial { \partial \mu   _ { X  _ { \alpha _ 1} }  }
  \cdots  
  T \frac \partial { \partial \mu   _ { X _  { \alpha  _ n } }    } 
  \tpf _ {  J  \bar J } ( x _ 1 - x _ 2 , \mu  ) 
  \right ]  _ { \mu_ X  = 0 }
  \label{dmu} 
\end{align}
with
\begin{align} 
\tpf _ {  J _ \alpha   \bar J _ \beta  } ( x , \mu  ) 
  \equiv
  Z
  ^ { -1 } 
  \;
  {\rm tr} \bigg \{ 
  \T J _ \alpha   ( x ) \bar J _ \beta   ( 0 )
  \exp \bigg [
    \frac 1T \bigg (
      \sum _ \gamma  \mu  _ { X _ \gamma  }
      X _ \gamma  - H _ {\rm SM }
      \bigg )
  \bigg ]
  \bigg \}
  \label{JJmu} 
  ,
\end{align}
where $ Z 
\equiv {\rm tr} \exp  \big[ \big( \sum _ \gamma 
        \mu  _ { X _ \gamma  } X _ \gamma  - H _ {\rm SM } \big) / T  \big] $
is the partition function at finite chemical potentials
of the slowly varying charges,
and $ H _ { \rm SM } $ is the Hamiltonian containing all Standard Model
interactions which are in equilibrium at the temperature of interest.
The traces in~\eqref{JJmu} and in $ Z $
run over states with definite
values of the conserved charges, which is why
only the slow charges appear in the exponential.
Note that one can introduce chemical
potentials for the $ X _ \gamma  $ only {\it  after} expanding in $ h $.
Therefore one cannot  write $ \Delta _ { f  f } ( \omega , \mu  ) $.
For vanishing $ h $,~(\ref{JJmu}) is 
diagonal in the active flavor indices, so that we can write
\begin{align} 
  \label{dia}
  \tpf _ {  J _ \alpha \bar J _ \beta } 
\equiv  \delta  _ { \alpha  \beta  } \tpf _ { \alpha  } 
   .
\end{align} 
The chemical potential of the operator $J_\alpha$ (in the sense of~\eqref{muA}) is 
$ \mu _ { J _ \alpha } = - \mu _ { X _ \alpha } $.
Therefore the function
$ e ^ { \mu _ { X _ \alpha } \tau }
\tpf _ {  \alpha } ( -i \tau , \vec x, \mu ) $ 
is anti-periodic in
$ \tau $, see~\eqref{period}, and
the Fourier decomposition of~(\ref{JJmu})
reads\footnote{See, e.g., chapter 8.1 of
               reference~\cite{Laine:2016hma}.}
\begin{align}
  \tpf _ { \alpha } ( -i \tau , \vec x, \mu )
  =
 T \widetilde { \sum \limits _ { p ^ 0 } }
  e ^ { - ( p ^ 0 + \mu _ { X _ \alpha } ) \tau }
  \tpf _ { \alpha } ( p ^ 0 , \vec x, \mu )
  \label{fouriermu}
   .
\end{align}
The Matsubara correlator on the right-hand side of~\eqref{fouriermu}
can be expressed through its spectral function%
\footnote{%
The spectral function for fermionic operators is defined
as in~\eqref{rho}, but with an anticommutator instead of the commutator.
}
via
\begin{align}
\tpf _ { \alpha } ( -i \tau , \mu )
=
\int 
   \frac { d \omega ' } { 2 \pi }
\;
e ^ { - \omega' \tau } \, \rho _ { \alpha } ( \omega ' , \mu )
\;
\Big\{
    \Theta ( \tau ) 
   &
       \big[ 1 - f _ { \rm F } (\omega' - \mu _ { X _ { \alpha } } ) \big]
     \nonumber \\
    &
    - { } 
\Theta ( - \tau ) f _ { \rm F } (\omega' - \mu _ { X _ { \alpha } } )
\Big\}
\label{tauspec}
.
\end{align}
The spectral function satisfies
\begin{align}
     \rho _ \alpha ( k, \mu ) 
   = \frac 1 { i } 
    \left [ 
   \tpf ^ { \rm ret } _ { \alpha } ( k , \mu ) -
     \tpf ^ { \rm adv } _ { \alpha } ( k , \mu )
   \right ] 
   \label{rhoal}
   ,
\end{align} 
and according to~\eqref{matsret},
the retarded and advanced correlators
are given by
\begin{align}
  \tpf _ { \alpha } ^ { \rm {ret, adv} } ( k , \mu )
  =
  \tpf _ { \alpha } \big( 
  k + u [- \mu _ { X _\alpha } \pm  i 0 ^ +]
  , \mu 
  \big)
  \label{anamu}
\end{align}
with real $ k ^ 0$, where $ u = ( 1, \vec 0 ) $ is the four-velocity of the 
plasma.

After summing over the Matsubara frequencies
we analytically continue $\omega $ towards the real 
axis which gives the $\order ( h ^ 2 ) $ contribution
to the retarded correlator 
\begin{align}
   \omega 
  \tpf _ { f ^ a _ { \vec{k} \lambda  } ( f ^ b _ { \vec p \lambda ' } 
   X _ { \alpha _ 1 }   \cdots X _ { \alpha  _ n } ) } 
   ^ { { \rm ret } } ( \omega )
    & { } 
  =
  - \; \delta  _ { \vec k \vec p } \, \delta  _ { \lambda \lambda ' }
    \sum \limits _ { \beta  \, ( i\, j \,l )}
  \frac{f ' _ { \text F } ( E  _ { \vec k i } ) } 
   { 2 E _ {\vec k i} }
   \bigg [  T \frac { \partial } { \partial \mu _ { X _ { \alpha _ 1} } } 
   \cdots 
         T \frac { \partial } { \partial \mu _ { X _ { \alpha _ n} } } 
   \label{retff}
  \\
  \nonumber
  \times 
  \Big\{ 
   h ^ \dagger _ {\beta  l }  T ^ b _ { l j } 
  T ^ a _ { j i } h  ^ { \phantom\dagger } _ {i \beta }  
  \,
    & \overline { u } _ { \vec k i \lambda  }
   \tpf ^{ \rm ret } _ { \beta  } ( k _ j, \mu   ) 
     u _ { \vec k l \lambda   }  
   -  \, 
   h ^ \dagger _ {\beta i } T ^ a _ {  i j }
  T ^ b _ { j l } h  ^ { \phantom\dagger } _ { l \beta }  
  \,
     \overline { u } _ { \vec k l \lambda  }
   \tpf ^{ \rm adv } _ {   \beta } ( k _ j, \mu   ) 
     u _ { \vec k i \lambda   }                  
  \\
 + \,
 h ^ \dagger _ {\beta  i } 
   T ^ a _ { l j }     T ^ b _ { j i } 
   h  ^ { \phantom\dagger } _ {l \beta }  
   \,
    &  \overline { v } _ { \vec k l \lambda  }  
   \tpf ^{ \rm ret } _ { \beta  } ( -k _ j , \mu  ) 
     v _ { \vec k i \lambda   }       
   -  \,
   h ^ \dagger _ {\beta  l } 
   T ^ b _ { i j } T ^ a _ { j l } 
   h  ^ { \phantom\dagger } _ { i \beta }  
   \,
      \overline { v } _ { \vec k i \lambda  }  
   \tpf ^{ \rm adv } _ {  \beta } ( -k _ j , \mu  ) 
     v _ { \vec k l \lambda   }  
   \Big\} 
   \bigg ] _ { \mu _X = 0 } 
   \nonumber
\end{align}
with $ k _ j \equiv ( E _ { \vec k j }, \vec k ) $.
Here and in the following we take $ \omega  \to 0 $  
on the right-hand side, as we did in equation~(\ref{retffM}).
Since
\begin{align}
\left[ \overline u _ { \vec k j \lambda }
 \tpf _ { J _  \alpha \bar J _ \beta } ^ { \text {ret} } ( q, \mu )
u _ { \vec k' i \lambda '}
\right] ^ *
=
\overline u _ { \vec k' i \lambda '}
 \tpf _ { J _ \beta \bar J _ \alpha } ^ { \text {adv} } ( q, \mu )
u _ { \vec k j \lambda }
\label{advret}
   ,
\end{align}
the right-hand side of~(\ref{retff}) is purely imaginary.

The computation of the correlator entering $ \gamma _ { f X } $ is analogous to the one
relevant for $ \gamma _ { f X X } $, with a few less extra steps.
In the computation of the latter we make use
of~(\ref{punkt1}), where the commutator vanishes, and obtain
the contribution
\begin{align}
\omega \,
{\rm Im \, }
\tpf ^ { \rm ret } _ { f ^ a _ { \vec k \lambda  } ( X _ \alpha X _ \beta ) } ( \omega )
=
-
{\rm Re }
\left[
\tpf ^ { \rm ret } _ { f ^ a _ { \vec k \lambda  } ( \dot X _ \alpha X _ \beta ) } ( \omega )
+
\tpf ^ { \rm ret } _ { f ^ a _ { \vec k \lambda } (  \dot X _ \beta X _ \alpha) } ( \omega )
+
\tpf ^ { \rm ret } _ { f ^ a _ { \vec k \lambda  } ( [X _ \alpha , \dot X _ \beta] ) } ( \omega )
\right]
\label{fXXp}
\end{align}
to the master formula~\eqref{gamma2}.
Since 
\begin{align} 
 \left[ \tpf ^ { \rm ret } _ { A B } ( \omega ) \right] ^ *
=
\tpf ^ { \rm ret } _ { A ^ \dagger  B ^ \dagger  } ( - \omega )
  \label{retABcc} 
\end{align} 
for bosonic operators $ A $ and  $ B $, 
the rightmost term in~\eqref{fXXp}
vanishes for  $ \omega \to 0 $, 
and we are left with the two terms in which the charge $ X $ 
without time derivative
appears to the right of  $ \dot X $. 
In these terms we can use the same line
of arguments as the one below~\eqref{fac} after having expanded the correlators to
quadratic order in $ h $,\footnote{In contrast to the computation of~\eqref{retff},
    here only one additional interaction is needed,
    because $ \dot X $ is already of order $ h $, see~\eqref{Xpunkt}.}
relating averages like
$ \langle \T J \bar J X \rangle $ to derivatives with respect to
chemical potentials of $ \langle \T J \bar J \rangle _ \mu $. The time derivative
of the charge is obtained from the Heisenberg equation of motion and reads
\begin{align}
  \dot X _ \alpha ( t ) = i 
   \sum \limits _ j    
   \int d ^ 3 x \left[
     \xbar N _ j ( t , \vec { x } ) h _ { j \alpha } J _ \alpha ( t , \vec x ) - \text {H.c.}
     \right]
  \label{Xpunkt}
  .
\end{align}
Using~\eqref{advret} and omitting terms of order $ \dM^2$, we obtain
(for $ n = 0,1$)
\begin{align}
  { \rm Re } \,
  \tpf ^ { \rm ret } _ { f ^ a _ { \vec k \lambda   } 
                        ( \dot X _ \alpha X _ \beta ^ n
                        ) 
                       } 
   ( \omega )
   =
   {} & -
    \sum \limits _ { ( i \, j )}
 \frac { 1 } { 4   E _ {\vec k i } }   
  \frac{f ' _ { \text F } ( E _ {\vec k i } ) }
       { f _ { \rm F } ( E _ {\vec k i } ) }
       \,
  T ^ a _ { i j }
  \bigg[\, \bigg( T \frac{ \partial } { \partial \mu _ { X _ \beta } } \bigg) ^ n
  \label{retfX} 
  \\
  \times \,
  \Big\{ 
  &
  \left ( 
     1 + e ^{ - \mu  _ { X _ \alpha  }/T }
  \right )
  f _ { \rm F } (  E _ {\vec k i } - \mu _ { X _  \alpha  } )
  h ^ { \dagger } _ { \alpha i }
  h ^ { \phantom \dagger } _ { j \alpha }
  \overline { u } _ { \vec k j \lambda  }  
\rho  _ \alpha  ( k _ i, \mu  ) 
  u _ { \vec k i \lambda   } 
  \nonumber
  \\
  -
  &
  \left ( 
     1 + e ^{ \mu  _ { X _ \alpha  }/T }
  \right ) 
  f _ { \rm F } ( E _ {\vec k i } + \mu _ { X _  \alpha  } ) 
  h ^ { \dagger } _ { \alpha j }
  h ^ { \phantom \dagger } _ { i \alpha }  
  \overline { v } _ { \vec k i \lambda  }  
\rho  _ \alpha  ( - k _ i, \mu  ) 
  v _ { \vec k j \lambda   } 
  \Big\}
  \bigg] _ { \mu _ X = 0 }
  \nonumber
  .
\end{align}

Analogously, we use~\eqref{punkt1} to relate 
$ \tpf ^ { \rm ret } _{ X ( f X \cdots )} $
to $ \tpf ^ { \rm ret } _ { \dot X ( f X \cdots ) } $, 
where the commutator vanishes once again. 
Using~\eqref{advret} and~\eqref{rhoal} we obtain the correlators
\begin{align}
  \omega
    \tpf _ { X _ \alpha ( f ^ a _ { \vec{k} \lambda   } 
   X _ { \beta _ 1 } \cdots
                               X _ { \beta _ n } ) } ^ { \text {ret} }
  ( \omega ) 
  =
  i \sum \limits _ {\gamma\,  ( i \, j )}
  &
   \frac{ f ' _ { \text F } ( E _ {\vec k i} ) } { 2 E _ {\vec k i} }
   \,
  T ^ a _ { i j }
  \,
  \bigg[ T \frac{ \partial } { \partial \mu _ { X _ {\beta _ 1} } } \cdots
         T \frac{ \partial } { \partial \mu _ { X _ {\beta _ n} } } 
    \label{retXf}  
   \\
   \times
  \Big\{ 
    &
    h ^ \dagger _ {\alpha  i } 
    h ^ { \phantom\dagger} _ { j \alpha }
    \overline { u } _ { \vec k j \lambda  } 
     \rho _ { \alpha } ( k _ i , \mu )
    u _ { \vec k i \lambda   }   
   {}-
   h ^ \dagger _ {\alpha  j } 
    h ^ { \phantom\dagger} _ { i \alpha }
    \overline { v } _ { \vec k i \lambda  } 
    \rho _ { \alpha } ( -k _ i , \mu )
    v _ { \vec k j \lambda   }    
    \Big\} 
    \bigg] _ { \mu _X = 0 }
    \nonumber
  ,
\end{align}
where once again we have dropped terms of order $\dM^2$.

The correlators containing only charges $ X $ are obtained similarly
to the steps that yield~\eqref{retfX}.
Equation~(\ref{punkt2}) gives
\begin{align}
  \tpf _ { X _ \alpha ( X _ \beta X _ \gamma ) } ^ { \text {ret} } ( \omega ) 
  =
  \frac { 1 } { \omega ^ 2 } 
  \left[ 
    \tpf _ { \dot X _ \alpha 
       ( X _ \beta X _ \gamma ) \mkern+2mu {\bm \dot { } } } ^ { \text {ret} } \,( \omega ) 
   + i 
   \big \langle [ X _ \alpha , ( X _ \beta X _ \gamma ) \mkern+1.5mu { \bm \dot { } }
   \, ]  
   \big \rangle 
   + \omega   
   \big \langle [ X _ \alpha , X _ \beta X _ \gamma ] 
   \big \rangle 
  \right]
  \label{XXX}
  .
\end{align}
The first commutator on the right-hand side drops out when
taking the imaginary part in~(\ref{gamma2}). The second one vanishes because the 
charges~(\ref{Xalpha}) commute. 
We rewrite the first term on the right-hand side as
\begin{align}
  \tpf _ { \dot X _ \alpha ( X _ \beta X _ \gamma )  \mkern+1.5mu { \bm \dot { } } }
       ^ { \rm ret } 
   \, 
   = 
   \tpf _ { \dot X _ \alpha ( \dot X _ \beta  X _ \gamma  ) } ^ { \rm ret } 
   + 
   \tpf _ { \dot X _ \alpha ( \dot X _ \gamma X _ \beta ) } ^ { \rm ret } 
   + 
   \tpf _ { \dot X _ \alpha ([ X _ \beta , \dot X _ \gamma ]) } ^ { \rm ret } 
  \label{XXXcomm}
  \,\, ,
\end{align}
where the third
term on the right-hand side of~\eqref{XXXcomm}
does not contribute to~\eqref{gamma2} due to the relation~\eqref{retABcc}.
Since   $ \dot X = \order ( h ) $,  at order $ h ^ 2 $ the first two terms can be obtained from
\begin{align}
   \tpf _ { \dot X _ \alpha ( \dot X _ \beta  X _ \gamma  ) } ^ { \rm ret } 
    ( \omega  ) 
   =
   \left [ 
   T \frac{ \partial } { \partial \mu _ { X _ {\gamma  } } } 
   \tpf _ { \dot X _ \alpha  \dot X _ \beta   } ^ { \rm ret } 
   ( \omega  , \mu  ) 
   \right ] _ { \mu  _ X = 0 }  
   \label{XpXpX} 
   .
\end{align} 
Then we take the thermodynamic limit replacing
$ \sum _ { \vec k } \to V \! \int _ { \vec k } $
and we find (the trace runs over spinor indices)
\begin{align}
  \frac 1 \omega 
  { \, \rm Im \, }
  \tpf _ { \dot X _ \alpha  \dot X _ \beta  } ^ { \text {ret} }
  ( \omega , \mu  )
  = { }
  - 
  \delta  _ { \alpha  \beta  } 
  \,
  V \! \!
  \int _ { \vec k }
  \sum \limits _ { i}
  |h ^ { \phantom \dagger}_ { i \alpha } | ^ 2 
  &
    \frac 1 { 4 E _ {\vec k i}  }    
  \frac {f ' _ { \text F } ( E _ {\vec k i} ) }
        {f _ { \text F } ( E _ {\vec k i} ) }
  \label{retXpXp}
  \\
  \times \,
  {\rm tr } 
  \Big \{
  \slashed k _ i
  \Big[
  &
  \left ( 
     1 + e ^{ - \mu  _ { X _ \alpha  }/T }
  \right ) 
  f _ { \rm F } ( E _ {\vec k i} - \mu _ { X _  \alpha  } ) 
     \rho   _ { \alpha } ( k _ i  , \mu )
  \nonumber
  \\
  + 
  &
  \left ( 
     1 + e ^{  \mu  _ { X _ \alpha  }/T }
  \right ) 
  f _ { \rm F } ( E _ {\vec k i} + \mu _ { X _  \alpha  } ) 
     \rho   _ { \alpha } ( -k _ i  , \mu )
  \Big]
  \Big \}  
  \nonumber
  .
\end{align}

\subsection{Kinetic equations}
\label{s:kineq}

In~\cite{Bodeker:2014hqa} the washout rate was written in
terms of charges. Here we express all rates in terms of the chemical
potentials $ \mu _ { X _ \alpha } $
by making use of
\begin{align}
\dX _ \alpha
=
\Xi _ { X _ \alpha X _ \beta } \frac { \mu _ { X _ { \beta } } } { T }
+
\frac 12
\Xi _ { X _ \alpha X _ \beta X _ \gamma }
\frac { \mu _ { X _ { \beta } } } { T }
\frac { \mu _ { X _ { \gamma } } } { T }
+ \order\big( \mu ^ 3 \big)
  \label{muXiX}
  .
\end{align}
When $ L _ { e \rm R } 	$ is slowly evolving,~\eqref{muXiX}
contains an additional term
$ \Xi
_ { X _ \alpha L _ { e \rm R } } ^ {\phantom \dagger}
\, \mu _ { L _ { e \rm R } } / T  $,
which does, however, not show up in the kinetic equations 
for $ f _ { \vec k \lambda  } $ and $ X _ \alpha  $, 
since
$ \sum _ \alpha  ( \Xi ^  { -1 } )
_ { X _ \beta X _ \alpha } \, \allowbreak
  \Xi
_ { X _ \alpha L _ { e \rm R } }
= \delta _ { X _ \beta L _ { e \rm R } } = 0 $. The same argument holds for
baryon number $ B $ in the temperature regime where its evolution is slow.
The $ \Xi _ { X X X } $ vanish, unless  there is some conserved charge which is nonzero.

In order to close the set of equations we need to specify the
susceptibilities in~(\ref{muXiX}). 
Furthermore, to evaluate the correlation
functions~(\ref{JJmu}) one has to switch from an  
ensemble in which the conserved charges have fixed values
to a grand-canonical one in which all charges  
$ Q _ a $ fluctuate, not
just the slowly varying ones.
Then  we have a relation similar
to~\eqref{muXiX}, but for \emph{all} charges,
\begin{align}
Q _ a
=
\sum \limits _ { b }
\chi ^ { \phantom \dagger } 
_ { a b } \frac { \mu _ { Q _ b } } { T }
+ \order\big( \mu ^ 3 \big)
  \label{muchiX}
  .
\end{align}
$\chi$ is the susceptibility matrix in the full grand-canonical ensemble, in which
the charges are odd functions of the chemical potentials. Therefore,
unlike in~\eqref{muXiX}, no terms of order $ \mu ^ 2 $ and no
equilibrium values appear 
in~\eqref{muchiX}. 

Combining relations~\eqref{muXiX} and~\eqref{muchiX} leads to
$ (\Xi ^ { -1 }) _ { a b }= (\chi ^ { -1 })_ {ab}  $, where $ a $ and  $b$
label only slow charges, and this way the equilibrium expectation values of the charges
can be obtained.
The matrix $ \chi $ depends on the temperature regime, 
see section~\ref{s:charges}.

From now on we understand the $ \tpf_ \alpha ( k, \mu ) $
to be defined in a grand-canonical description,
where the $ \mu $ label chemical potentials associated with
\emph{all} charges.

We expand our kinetic equations to quadratic order in
slowly varying chemical potentials,
then the higher order terms in~\eqref{muXiX} do not
contribute.
The term with $ \Xi _{ XXX }$ on the right-hand side of~\eqref{muXiX} cancels the
second term in square brackets in~\eqref{g3}.
The corresponding susceptibility of the occupancies
$ \Xi _ { f f f } $, which appears only in~\eqref{gamma2},
leads to cancellation of the coefficient $ \gamma _ { f f f }$,
see appendix~\ref{a:gfff}, and $ \Xi _ { f f f f } $ in~\eqref{gamma3} does
the same with the coefficient $ \gamma _ { f f f f } $.
At order $ h ^ 0 $ all other $ \Xi _ { a b c } $ vanish identically, and
the only other nonzero $ \Xi _{ a b c d } $ are those with four charges $ \dX $,
which, however, enter~\eqref{gamma3} only for a coefficient multiplying three
factors of $\dX$ in~\eqref{eom}, which is beyond our expansion to order~$ \mu ^ 2 $.

The susceptibilities of the sterile-neutrino occupancy read
(without sum over~$ \vec k $ or~$ \lambda  $)
\begin{align}
  \Xi
       _ { f ^ a _ { \vec k \lambda } f ^ b _ { \vec k \lambda  } }
  =
  T ^ a _ { i j }
  T ^ b _ { j i }
  \,
  f _ { \text F } ( E _ {\vec k i} ) 
  [ 1 - f _ { \text F } ( E _ {\vec k j} ) ] 
  \label{chiff}
  \,
  .
\end{align}
Plugging~\eqref{retffM}, \eqref{retff}, and~\eqref{retfX} 
into the respective master
formulae~\eqref{kubo}--\eqref{gamma3}
we obtain the kinetic equations 
\begin{align}
\nonumber 
  ( \dot f _ { \vec k \lambda  } ) _ { m n }  
  =
    \frac { i } { 2 E _ { \vec k m } }
  \Bigg\{
  &
  \dM ^ 2 _ { m n }  
  \,
  (f _ { \vec k \lambda } ) _ { m n }
  \\
  + \sum \limits _ {\alpha \, l }
   \bigg [ 
     \overline { u } _ { \vec k l \lambda   }
     \Big(
    &
     h  _ { n \alpha }
     \tpf ^{ \rm ret } _ { \alpha  } ( k _ l , \mu   ) 
     h ^ { \dagger } _ { \alpha l }
     \big[
        ( f _ { \vec k  \lambda } ) _ { m l } 
        -
        \delta _ { m l }
        f _ { \rm F } ( E _ { \vec k l } - \mu _ { X _ \alpha } )
     \big]
  \nonumber
  \\
     \vphantom{  \sum \limits _ {\alpha  \, l} \Big\{ }
   - { }   &
     h  _ { l \alpha }
     \tpf ^{ \rm adv } _ { \alpha  } ( k _ l , \mu   ) 
     h ^ { \dagger } _ { \alpha m }
     \big[
        ( f _ { \vec k \lambda  } ) _ { l n } 
        -
        \delta _ { l n }
        f _ { \rm F } ( E _ { \vec k l } - \mu _ { X _ \alpha } )
     \big]
     \Big)
     u _ { \vec k  l \lambda   }
  \nonumber
  \\
     \vphantom{  \sum \limits _ {\alpha  \, l} \Big\{ }
   +  \overline { v } _ { \vec k  l \lambda   }
     \Big(
     &
     h  _ { m \alpha }
     \tpf ^{ \rm ret } _ { \alpha  } ( - k _ l, \mu   ) 
     h ^ { \dagger } _ { \alpha l }
     \big[
        ( f _ { \vec k  \lambda } ) _ { l n } 
        -
        \delta _ { l n }
        f _ { \rm F } ( E _ { \vec k l } + \mu _ { X _ \alpha } )
     \big]
  \nonumber
  \\
   \vphantom{  \sum \limits _ {\alpha  \, l} \Big\{ }
      - { } &
     h  _ { l \alpha }
     \tpf ^{ \rm adv } _ { \alpha  } ( - k _ l , \mu   ) 
     h ^ { \dagger } _ { \alpha n }
     \big[
        ( f _ { \vec k \lambda  } ) _ { m l } 
        -
        \delta _ { m l }
        f _ { \rm F } ( E _ { \vec k l } + \mu _ { X _ \alpha } )
     \big]
     \Big)      v _ { \vec k l \lambda   }
  \bigg ] 
  \Bigg\}
  \nonumber
  \\
  { } 
  + \; \order \big( \mu ^ 3 ,  h ^ 2 & \dM^2, \dM^4, h^4 \big)
  \label{fkineq}
  ,
\end{align}
for those elements of the 
occupancy matrix for which 
$ | \dM ^ 2 _ { mn } | /T \ll \omuv $
(including, of course, the diagonal elements).
For the other elements the right-hand side vanishes in our approximation.
The sum is over indices $ l $ for which 
$ | \dM ^ 2 _ { ml } | /T \ll \omuv $.
$ f _ { \vec k + }  $ does not appear on the right-hand side of the
kinetic equation for $ f_ { \vec k - }  $, and vice versa.
 
In equation~\eqref{retXpXp} we express $ \slashed k $ through 
the completeness
relation of the $ u $ or $ v $ spinors.
Together with equation~\eqref{retXf}, and taking the limit $ V \to \infty$,
we obtain the kinetic equation for the charge density
$ n _ { X _ \alpha } \equiv X _ \alpha / V $
\begin{align}
\dot n _ { X _ \alpha }
= 
  \sum \limits _ { ( i\, j ) \, \lambda }
    \int _ { \vec k }
\frac { 1 } { 2 E _ { \vec k  i } }
  \Big\{
     &
     \overline { u } _ { \vec k i \lambda}
     h ^ {\phantom \dagger } _ { i \alpha }
     \rho _ { \alpha } ( k _ i , \mu )
     h ^ \dagger_{ \alpha j }
     u _ { \vec k i \lambda }  
     \,
     \big[
     ( f _ { \vec k \lambda } ) _ {  i j }
     -
     \delta _ { i j }
     f _ { \rm F } ( E _ { \vec k  i } - \mu _ { X _ \alpha } )
     \big]
     \nonumber
     \\ \phantom {   \sum \limits _ { ( i\, j ) \, \vec k } }
     - \,
     &
     \overline { v } _ { \vec k i \lambda}
     h ^ {\phantom \dagger } _ { j \alpha }
     \rho _ { \alpha } ( - k _ i , \mu )
     h ^ \dagger_{ \alpha i }
     v _ { \vec k i \lambda }  
     \,
     \big[
     ( f _ { \vec k \lambda } ) _ { i j }
     -
     \delta _ { i j }
     f _ { \rm F } ( E _ { \vec k  i } + \mu _ { X _ \alpha } )
     \big]
     \nonumber
 \Big\}
 \nonumber 
 \\ \phantom {   \Big\{ } 
 + \;  \order \big(  \mu ^ 3, h ^ 2  & \dM^2, h^4 \big)
 \label{Xkineq}
 .
\end{align}
When additional processes are slow, one needs additional kinetic equations.
Around $ T \sim $ 130 GeV, this is the case for the $ B+L $ violating 
electroweak sphaleron processes~\cite{DOnofrio:2014rug}.
Then it is convenient to include
a kinetic equation for $ B $\cite{Khlebnikov:1996vj,Burnier:2005hp}, since
$ B $ is not violated by the sterile-neutrino Yukawa interaction
so that only the sphaleron rate enters this equation.  When $ T \sim $
85 TeV, the lepton number carried by right-handed electrons
evolves slowly, and one has to include the
kinetic equation for $ L _ { e  { \rm R  } } $~\cite{Bodeker:2019ajh}.

Using a different approach from ours, 
an equation  similar to~(\ref{fkineq}) was derived previously
in~\cite{Ghiglieri:2017gjz},\footnote{See equation~(2.29)
         of reference~\cite{Ghiglieri:2017gjz}.}
assuming $ M _ i \ll T $ for all Majorana
masses, so that the condition~(\ref{degen})  is satisfied.
In~\cite{Ghiglieri:2017gjz} the chemical potential
for $ L _ \alpha  $ appears in $ f _ { \rm F } $  
instead of the one for $  { X _ \alpha  } $, which is nevertheless
consistent, see section~\ref{s:charges}.
Unlike the equation in~\cite{Ghiglieri:2017gjz}, 
our~(\ref{fkineq}) contains not only scattering contributions, but
also dispersive ones, see section~\ref{s:muosc}.
In~\cite{Ghiglieri:2017gjz} the latter are incorporated at a later stage. 
Aside from that, the first term in the curly bracket in~(\ref{fkineq}) 
(which contains the $ u $-spinors) 
is equivalent to the corresponding one
in~\cite{Ghiglieri:2017gjz}.%
\footnote{%
  The helicity diagonal contribution
  containing the $ v $-spinors in~\cite{Ghiglieri:2017gjz}
is not consistent with our equation~(\ref{fkineq}).
However, it becomes consistent after
applying (2.22) of~\cite{Ghiglieri:2017gjz}.
\label{replace}
 }
For vanishing Majorana masses~(\ref{Xkineq}) coincides with
the corresponding equation in~\cite{Ghiglieri:2017gjz}.%
\footnote{See equation (2.32) of reference~\cite{Ghiglieri:2017gjz}.}
Furthermore, for non-vanishing Majorana masses, our 
contributions containing the $u$-spinors also appear there. The $ v $-spinor
contribution is equivalent, after the replacement described in 
footnote~\ref{replace}. 

In reference~\cite{Ghiglieri:2015jua} kinetic equations for
the spin-averaged occupancies of  
sterile neutrinos without near mass-degeneracy and for 
lepton numbers  in the Higgs phase have been obtained. 
There the spin-asymmetry of the sterile neutrinos has been neglected.%
\footnote{See equations~(2.21) and~(2.24) of
reference~\cite{Ghiglieri:2015jua}.} 
In reference~\cite{Ghiglieri:2019kbw} the same authors have obtained 
equations for
a hierarchical system with one light and two heavy sterile neutrinos in
the Higgs phase. 
There the kinetic equations are given in terms of the 
retarded correlator of $ J $
as a function of slowly varying chemical potentials, like in our~\eqref{fkineq}.
The terms multiplying these correlators are given to linear order
in slowly varying quantities.
Using the relation%
\footnote{Since we have made the transition to the grand-canonical
          description, $ \mu $ in equation~\eqref{uv}
          now denotes the chemical potentials of all charges.}
\begin{align}
 \overline v _ { \vec k i \lambda }
\tpf ^ { \rm ret } _ { \alpha  } ( - q , \mu )
v _ { \vec k i \lambda }
\;
=
{ }
-
\overline u _ { \vec k i ,-\lambda }
\tpf ^ { \rm adv } _ { \alpha  } ( q , - \mu )
u _ { \vec k i, -\lambda  }
\label{uv}
,
\end{align}
(no sum over repeated indices)
which is valid when the Standard-Model CP violation is neglected, 
together with~\eqref{advret}, we can reproduce the kinetic
equations for the light flavor 
and the heavy ones,\footnote{See equations~(2.5)
and~(2.6) of reference~\cite{Ghiglieri:2019kbw}.}
as well as the one for the lepton asymmetries.%
\footnote{See equation~(2.4) of reference~\cite{Ghiglieri:2019kbw}.}
 
\subsection{Small Majorana masses}
\label{s:lepto}

In low-scale leptogenesis~\cite{Akhmedov:1998qx,Asaka:2005pn} 
the sterile-neutrino masses are small compared to the temperature,
so that typically 
$ M _ i \ll | \vec k |$. 
Then they can be neglected in the terms containing their
(also small) Yukawa couplings, so that  
the helicity eigenspinors $ u $ and $ v $ are purely 
right- and left-handed. Since the operator $ J $ 
is purely 
left-handed, the terms with $ v _ + $ or $ u _ - $ drop out.
Then we also have
$  u _ { \vec k i + } \overline u _ { \vec k i + } =
   v _ { \vec k i - } \overline v _ { \vec k i - } =
  {\rm P } _ { \rm R }  \slashed k  $ with the chiral
projector $ { \rm P} _ { \rm R } \equiv ( 1 + \gamma ^ 5 ) / 2$.
Furthermore, condition~\eqref{degen} is trivially satisfied,
so that the kinetic equations simplify to
\begin{align}
  ( \dot f _ { \vec k +  } ) _ { m n }  
  =
    \frac { i } { 2 | \vec k | }
  \bigg \{
  \big[ M ^ 2,
   &
f _ { \vec k +  } \big] _ { m n } 
  \nonumber \\ {}
  + 
  \sum \limits _ {\alpha \, l }
    {\rm tr } \Big [  \slashed k
     \Big(
     &
     h  _ { n \alpha }
     \tpf ^{ \rm ret } _ { \alpha  } ( k , \mu   ) 
     h ^ { \dagger } _ { \alpha l }
     \big[
        ( f _ { \vec k  + } ) _ { m l } 
        -
        \delta _ { m l }
        f _ { \rm F } ( | \vec k | - \mu _ { X _ \alpha } )
     \big]
  \nonumber
  \\
     \vphantom{  \sum \limits _ {\alpha  \, l} \Big\{ }
   - { }   &
     h  _ { l \alpha }
     \tpf ^{ \rm adv } _ { \alpha  } ( k , \mu   ) 
     h ^ { \dagger } _ { \alpha m }
     \big[
        ( f _ { \vec k +  } ) _ { l n } 
        -
        \delta _ { l n }
        f _ { \rm F } ( | \vec k | - \mu _ { X _ \alpha } )
     \big]
     \Big)
     \Big ]
   \bigg \} 
  \label{fplep}
  ,
\\[8pt]
  ( \dot f _ { \vec k -  } ) _ { m n }  
  =
    \frac { i } { 2 | \vec k | }
  \bigg \{ 
  \big[ M ^ 2, 
  &
   f _ { \vec k -  } \big] _ { m n } 
  \nonumber \\
  {} + 
  \sum \limits _ {\alpha \, l }
    {\rm tr } \Big [  \slashed k
     \Big(
     &
     h  _ { m \alpha }
     \tpf ^{ \rm ret } _ { \alpha  } ( - k , \mu   ) 
     h ^ { \dagger } _ { \alpha l }
     \big[
        ( f _ { \vec k - } ) _ { l n } 
        -
        \delta _ { l n }
        f _ { \rm F } ( | \vec k | + \mu _ { X _ \alpha } )
     \big]
  \nonumber
  \\
   \vphantom{  \sum \limits _ {\alpha  \, l} \Big\{ }
      - { } &
     h  _ { l \alpha }
     \tpf ^{ \rm adv } _ { \alpha  } ( - k , \mu   ) 
     h ^ { \dagger } _ { \alpha n }
     \big[
        ( f _ { \vec k -  } ) _ { m l } 
        -
        \delta _ { m l }
        f _ { \rm F } ( | \vec k | + \mu _ { X _ \alpha } )
     \big]
     \Big)
  \Big ]
  \bigg \} 
  \label{fmlep}
  ,
\end{align}
and (again $ n _ {X _ \alpha } \equiv X _ \alpha / V $)
\begin{align}
\dot n _ { X _ \alpha }
=
  \sum \limits _ { ( i\, j ) } 
  \int _ { \vec k }
\frac { 1 } { 2 | \vec k  | }
  {\rm tr } \Big[ \slashed k
  \Big\{
     &
     h ^ {\phantom \dagger } _ { i \alpha }
     \rho _ { \alpha } ( k  , \mu )
     h ^ \dagger_{ \alpha j }
     \,
     \big[
     ( f _ { \vec k + } ) _ {  i j }
     -
     \delta _ { i j }
     f _ { \rm F } ( | \vec k | - \mu _ { X _ \alpha } )
     \big]
     \nonumber
     \\ \phantom {   \sum \limits _ { ( i\, j ) \, \vec k } }
     - \,
     &
     h ^ {\phantom \dagger } _ { j \alpha }
     \rho _ { \alpha } ( - k  , \mu )
     h ^ \dagger_{ \alpha i }
     \,
     \big[
     ( f _ { \vec k - } ) _ {  i j }
     -
     \delta _ { i j }
     f _ { \rm F } ( | \vec k | + \mu _ { X _ \alpha } )
     \big]
 \Big\} \Big]
 .
 \label{Xlepto}
\end{align}
In~\eqref{fplep}--\eqref{Xlepto} we have $ k ^ 0 =  | \vec k | $, 
the traces refer to spinor indices, and we have neglected terms of
order $ \mu  ^ 3 $, as well as terms of order $  h ^ 2   M ^ 2 $, $   M^4
$ and $  h^4 $.
Similar equations have been obtained in reference~\cite{Hernandez:2016kel}.
Keeping in mind that they use the index convention of 
reference~\cite{Sigl:1992fn},
we can reproduce their kinetic equation\footnote{See
            equation~(2.14) of reference~\cite{Hernandez:2016kel}.}
for the sterile neutrino occupancies by 
setting $  { \rm tr } 
  \big [ \slashed k  \tpf _ { \alpha } ^ { \rm ret } ( k , \mu ) \big]
  = { \rm tr } 
  \big [ \slashed k  \tpf _ { \alpha } ^ { \rm adv } ( k , \mu ) \big] ^ \ast
  \to
  - T ^ 2 / 4 + i | \vec k | \left( \gamma ^ { ( 0 ) } + \mu _ \alpha 
                                                         \gamma ^ { ( 2 ) } \right)$
in our~\eqref{fplep} and~\eqref{fmlep}, and neglecting terms quadratic 
in chemical potentials. 
Recalling~\eqref{rhoal}, we can also reproduce
their kinetic equation\footnote{See equation~(2.18) of reference~\cite{Hernandez:2016kel}.}
corresponding to our~\eqref{Xlepto} in the same way.
Setting $ \mu  = 0 $ in the spectral function $ \rho _ \alpha $ 
in~(\ref{Xlepto}) and
expanding the Fermi distribution to linear order in $ \mu  _ { X _ \alpha  }
$, one obtains the washout term which was found in~\cite{Bodeker:2014hqa}.

\section{Susceptibilities and right-handed electron number}  
\label{s:charges}

The computation of the susceptibilities in (\ref{muchiX}) 
in the symmetric phase 
is described in~\cite{Bodeker:2014hqa}, where
the leading order (in Standard Model couplings) 
contributions to the pressure have been obtained.
There also the $ \order ( g ) $
corrections and some of the $ \order( g^2) $ contributions can be found,
and the remaining $ \order(g^2) $ contributions have been
obtained in~\cite{Bodeker:2015zda}.
In the symmetric phase, the only  conserved charge  that is correlated 
with the $ X _ \alpha $ at all temperatures is the U(1)-hypercharge~$ Y $.
The zero-momentum mode of the hypercharge gauge field plays the role of
the corresponding chemical potential $ \mu _ Y $ in the path integral
formalism~\cite{Khlebnikov:1996vj} where it ensures
hypercharge neutrality of the plasma.

When electroweak sphalerons are in equilibrium,
our statistical operator which determines $ \tpf _ \alpha  $ contains
$ \mu _ { X _ \alpha  } X _ \alpha $,  
but no separate chemical potential for baryon number because
the latter is not conserved. 
In reference~\cite{Ghiglieri:2017gjz}
$ \mu _ \alpha L _ \alpha + \mu _ B B $ appears.
However, using the equilibrium conditions one 
can match the coefficients
which gives $   \mu _ \alpha  = \mu _ { X _ \alpha }$, and
$ \mu _ B = - \sum _ \alpha \mu _ { X _ \alpha } / 3 $. 
Therefore the chemical potentials appearing in the distribution functions in
our kinetic equations are consistent with those in~\cite{Ghiglieri:2017gjz}.

In the temperature range 
$ 8.5\cdot 10^4\text{ GeV} \ll T \ll 10^9\text{ GeV}$ 
the lepton number $ L _ { e \rm R } $
carried by right-handed electrons
is not yet efficiently violated by the electron Yukawa
coupling~\cite{Bodeker:2019ajh} so that it constitutes an additional conserved
charge, and
we have to introduce a corresponding chemical potential
in~(\ref{muchiX}).
The relation between all charges and their chemical potentials
has been obtained in~\cite{Bodeker:2019ajh}.\footnote{See equation (A.2)
   of reference~\cite{Bodeker:2019ajh}.}
Hypercharge neutrality implies
\begin{align}
\mu _ Y &= \frac {1}{11} \mu _ { L _ {e\rm R } } +
        \frac {8}{33} \sum\limits_\alpha \mu_ { X _ \alpha }
  \label{muY}
  ,
\end{align}
and the relation between $ L_ { e \rm R } $ and its chemical potential
reads
\begin{align}
\mu _ { L _  { e \rm R } }
&=
-\frac 56 \mu _ { X _ e }
+ \frac {4}{15} \left(  \mu_ { X _ \mu } + \mu_ { X _ \tau } \right)
+ \frac {33}{5 \, T ^ 2 V }    L _ { e \rm R  }  
  \label{muL}
  .
\end{align}

Around $ T \sim 8.5 \cdot 10^4\text{ GeV} $
the interactions mediated by the electron Yukawa coupling happen
at a rate comparable to the Hubble rate, so that $ L _ { e \rm R } $ is slowly
varying according to the evolution equation in reference~\cite{Bodeker:2019ajh}.
The relation between charges and chemical potentials remains the
same as above, and also~\eqref{muY} and~\eqref{muL} are valid.
In particular, even though the sterile-neutrino interactions
do not violate the conservation of $ \LeR $,
and the electron Yukawa coupling does not violate the one of $ X _ \alpha $,
the evolution equations of $ \LeR $ and the $ X _ \alpha $
are coupled through the matrix of susceptibilities.

At much lower temperatures the right-handed electron
lepton number is in equilibrium, and no chemical potential $ \mu _ { \LeR} $ 
is included. Between $ 160\text{ GeV} \ll T \ll 8.5\cdot 10^4\text{ GeV} $,
all Standard Model interactions are in equilibrium, and the relation between
charges and chemical potentials is found in~\cite{Bodeker:2014hqa}.%
\footnote{See equation~(63) of reference~\cite{Bodeker:2014hqa}.} Imposing
hypercharge neutrality leads to the relation $ \mu _ Y =
\mu _ Y ( \mu _ { X _ \alpha } ) $ given in~\cite{Ghiglieri:2017gjz}.%
\footnote{See equation~(4.2) of reference~\cite{Ghiglieri:2017gjz}.}

Around $ T \sim $ 130 GeV 
electroweak sphalerons freeze out~\cite{DOnofrio:2014rug}, so that baryon number
is a slow variable, and we have to introduce
a corresponding chemical potential $ \mu _ B $. 
The relation between charges and chemical potentials can be
found in~\cite{Eijima:2017cxr}, where also the developing Higgs 
expectation value is taken into account.\footnote{See equations~(3.8) and~(3.9)
                                                  of reference~\cite{Eijima:2017cxr}.}
Hypercharge neutrality yields the relation
$ \mu _ Y ( \mu _ B, \mu _ { X _ \alpha } ) $ found in~\cite{Ghiglieri:2017gjz}.%
\footnote{See equations~(4.5) and~(4.6) of reference~\cite{Ghiglieri:2017gjz}.}
In this temperature regime our statistical operator contains
$ \mu _ { X _ \alpha } X _ \alpha + \mu _ B B $, while the one in
reference~\cite{Ghiglieri:2017gjz} is the same as in the high-temperature regime
discussed above. Matching the chemical potentials again yields
$  \mu _ \alpha = \mu _ { X _ \alpha }  $, and this time
$ \mu _ B ^ {\rm there} = \mu _ B ^ { \rm here}
                          - \sum _ \alpha \mu _ { X _ \alpha } / 3 $, so that again
the chemical potentials in the distribution functions appearing in the kinetic
equations coincide.

Deep in the Higgs phase (${T \ll 130\text{ GeV}}$) 
the susceptibilities have a non-trivial
dependence on the temperature.
They have been studied
in~\cite{Ghiglieri:2018wbs}.%
\footnote{See appendix A of reference~\cite{Ghiglieri:2018wbs}.}

\section{Standard Model correlators in the sym\-metric phase}
\label{s:sm}

Deep in the symmetric phase one has to distinguish two temperature
regimes. When $ { M _ i \gg g T } $, 
at leading order in the Standard Model couplings
the dissipative (imaginary) part of
$ \tpf ^ { \rm ret} _ \alpha  $ is determined by  
hard $ 2 \leftrightarrow  2 $ scattering processes. 
For $ M _ i \lsim g T $, nearly collinear $ 1 \leftrightarrow  2 $ decays 
and inverse decays involving a Higgs boson, a
SM lepton and a sterile neutrino, plus the same process
with additional soft scatterings, which are sometimes referred to as
$ 1 n \leftrightarrow  2 n $ processes, also contribute at
leading order~\cite{Anisimov:2010gy}. 

The multiple soft scatterings need to be resummed, which is referred
to as Landau-Pomeranchuk-Migdal (LPM) 
\cite{Landau:1953um,Landau:1953gr,Migdal:1956tc}
resummation. 
The result gives an imaginary contribution to $ \tpf ^ { \rm ret} _ \alpha  ( k _ j , \mu ) $
which can be computed along the lines 
of~\cite{Anisimov:2010gy,Ghiglieri:2016xye}.
It can be expressed in terms of the 2-component vector function
$ \vec f _ j( \vec b  ) $ 
and the  scalar function
$ \psi _ j ( \vec b ) $, which both
depend on the 2-component impact parameter vector $ \vec b $.
They can be obtained by solving the differential equations 
in~\cite{Anisimov:2010gy} with $ M _ N $ replaced by $ M _ j $.
It is straightforward to generalize the analysis of~\cite{Anisimov:2010gy}
to non-zero chemical potentials
which gives
\begin{align}
   \tpf ^ { \rm ret,\,LPM } _ \alpha  
   ( k _ j, \mu  )  
   & 
   = \frac i  2
   \int 
   \frac { d p _ \parallel } { 2 \pi  } 
   \frac 1 { | \vec k | - p _ \parallel } 
   \left   [ f _ { \rm B } \left 
   ( p _ \parallel - | \vec k | + \frac { \mu  _ Y } 2 \right ) 
     + f _ { \rm F } \left ( p _ \parallel - \mu  _ { X _ \alpha  } 
         + \frac { \mu  _ Y }  2 \right ) 
   \right ]  
  \nonumber \\
    &
   \times
   { \rm P }  _ { \rm L }  
  \lim _ { \vec b \to \vec 0 } 
   \bigg \{
      \left ( \gamma  ^ 0 - \vec { \hat  k } \cdot \gvec \gamma  \right )
   { \rm Re} \, \psi _ j ( \vec b )
   + 
   \frac 1    { 8 p _ \parallel ^ 2 } 
   \left ( \gamma  ^ 0 + \vec { \hat  k } \cdot \gvec \gamma  \right )
   { \rm Im} \, \nabla _ { \vec b } \! \cdot \vec f _ j  ( \vec b )
   \bigg \} 
   \label{lpm}
   .
\end{align} 
Here $ f _ { \rm B } (E) \equiv 1/(e^{E/T}-1)$ is
the Bose-Einstein distribution,
$ { \rm P }  _ { \rm L }  \equiv ( 1 - \gamma  _ 5 ) /2 $ is a 
chiral projector, and $ \vec { \hat  k } \equiv \vec k /| \vec k | $.
The second term  in the curly bracket 
is of order $ g ^ 2 $ times the first. 
Nevertheless, it  
has to be kept because when sandwiched between the $ u $- and 
$ v $-spinors, the first term gets multiplied by $ M _ j ^ 2 $ which here is
assumed to be order   $  g ^ 2 T ^ 2 $ or smaller. 
The LPM contribution was computed in~\cite{Ghiglieri:2017gjz}, 
where the result does not contain a chiral projector.%
\footnote{See equation~(3.4) of reference~\cite{Ghiglieri:2017gjz}.}
Otherwise it is consistent with equation~(\ref{lpm}) 
(the second term in the curly
bracket differs from the corresponding one in~\cite{Ghiglieri:2017gjz}
only by higher orders in~$ g $).
Chiral projectors were correctly included in~\cite{Ghiglieri:2017gjz}
when the result was sandwiched between the $ u $ 
and $ v $ spinors which makes it consistent with ours 
(cf.\ footnote~\ref{replace} on page \pageref{replace}).
The $ 2\to 2 $ scattering contributions to the rate coefficients
have been computed in~\cite{Ghiglieri:2017gjz}.

\subsection{Dispersive contributions} 
\label{s:muosc}

The imaginary parts of the 2-point functions in~(\ref{fplep}),~\eqref{fmlep}
have been computed in~\cite{Ghiglieri:2017gjz} at nonzero chemical potentials. 
Here we compute the real part 
in the symmetric phase 
which modifies the dispersion relations 
of the sterile neutrinos.
We include the chemical potentials to linear
order and we work at leading order in Standard Model couplings,
assuming $ M _ i \ll  | \vec k | $.
The leading order is contained in the 1-loop contribution, 
which reads 
\begin{align}
  \tpf _ { \alpha } ( k ^ 0, \vec k , \mu )
  =
  T \widetilde { \sum \limits _ { p ^ 0 } } \int _ { \vec p }
  \frac { 2 \, {\rm P} _ {\rm L}  
   ( \slashed p - \mu _ { \ell _ \alpha } \slashed u ) }
        { ( p - \mu _ { \ell _ \alpha } u ) ^ 2 \,
          [ (k - p - \mu _ { \varphi }\, u ) ^ 2 - m _ \varphi ^ 2 ] }
  \label{JbarJ}
\end{align}
in imaginary-time. 
The factor $ 2 $ is the dimension of the representation of the weak SU(2). 
The chemical potentials in~\eqref{JbarJ} are the ones carried 
by the field operators, cf.~(\ref{muA}).
They have the opposite sign compared to the chemical potentials
carried by the particles which they annihilate.
Note that in~\eqref{JbarJ}  
\begin{align}
\mu _ { \varphi }
 = - \frac {\mu_ Y } { 2} 
\label{muhiggs}
,
\end{align}
appears, rather than $ \mu  _ { \widetilde \varphi  } $.%
\footnote{Equation~(\ref{muhiggs}) is consistent with~\cite{Ghiglieri:2017gjz}
and~\cite{Bodeker:2019ajh},
where the Higgs chemical potential is defined as a particle chemical
potential.} 
The relation between the hypercharge
chemical potential $ \mu _ Y $ and the chemical potentials of
the slowly varying charges depends on the temperature,
see section~\ref{s:charges}.

The leading contribution from~(\ref{JbarJ}) is due to 
soft Higgs momenta, which are cut off 
by the thermal 
Higgs mass~\cite{Weldon:1982bn}
\begin{align}
  m _ \varphi ^ 2 = \frac { 1 } { 16 } ( 3 g ^ 2 + { g ' } ^ 2 + 4 h _ t ^ 2 + 8 \lambda )
  \left( T ^ 2 - T _ 0 ^ 2 \right)
  \label{mphi}
\end{align}
in the Higgs propagator, which gives rise to an infrared enhancement. 
Here $ T _ 0 = 160\text{ GeV}$, and 
$ g $ and $ g ' $ are the weak SU(2) and hypercharge U(1) gauge couplings,
respectively.
Our normalization is such that covariant derivatives are
  $ D _ \mu  =
  \partial _ \mu  + i y _ \alpha g ' B _ \mu  + \cdots $
with the hypercharge gauge field $ B $,
and $ y _ \varphi = 1/2 $ for the Higgs field $ \varphi  $.
Furthermore,  $ h _ t $ is the top quark Yukawa coupling, and 
the quartic term in the Higgs potential  is given by
$ \lambda ( \varphi ^\dagger \varphi ) ^ 2 $.

After summing over the imaginary fermionic frequency 
$ p ^ 0 $ we analytically continue to real $  k ^ 0 $
according to~\eqref{anamu}, and obtain
\begin{align}
  { \rm \,Re\, }   { \rm tr } 
  \big [ \slashed k  \tpf _ { \alpha } ^ { \rm ret } ( k , \mu ) \big]
  = 
  { }
  - \frac { T ^ 2 } { 4 }
  + \frac {  m _ \varphi T } { 4 \pi | \vec k | } \mu _ { \varphi }
  + \order ( g^2 \mu, \, \mu ^ 2 )
  \label{Re}
  ,
\end{align}
with $ k ^ 0 = | \vec k | $. The $ g $ in the higher order terms in~\eqref{Re}
stands for a generic Standard Model coupling.
The  first term on the right-hand side of~(\ref{Re}) gives 
rise to the thermal mass. 
The $ \mu  $-dependent contribution  is not simply a correction
of the thermal mass, but it depends on momentum. 
In particular, it is enhanced at small $ \vec k $.
The leading correction from chemical potentials is independent of
$ \alpha $. In reference~\cite{Ghiglieri:2018wbs} an expression corresponding
to~\eqref{Re} in the broken phase has been obtained.\footnote{%
See equation~(5.7) of reference~\cite{Ghiglieri:2018wbs}.}

\section{Summary}
\label{s:sum} 

In this paper we have obtained non-linear kinetic
equations which describe the time evolution of 
sterile-neutrino phase space densities and charge densities
carried by Standard Model particles
by generalizing the approach of reference~\cite{Bodeker:2014hqa} 
to  include non-linear terms.
To determine the coefficients in these equations we have
matched not only real-time two-point functions  in the
effective kinetic equations for thermal fluctuations to those
in thermal field theory, like
in~\cite{Bodeker:2014hqa}, but also higher point functions. 
The sterile neutrinos have been integrated out using a  path integral over their
Fourier coefficients, which correspond to their creation and annihilation
operators. We have included only the leading order in their Yukawa coupling
and in their  Majorana mass squared differences. 
This way we have obtained relations
between the rate coefficients and real-time correlation functions
of Standard Model fields, evaluated at finite temperature and chemical
potentials for charges which are conserved or slowly violated (in
the case of $\LeR$ or $B$ for certain temperatures)
by the Standard Model interactions.
The rate coefficients are infrared safe in the sense that they are
well behaved
when a parameter characterizing a slow interaction vanishes. 

The kinetic equations and the relations for the rate coefficients found
in this paper
are mostly consistent with the ones obtained
in reference~\cite{Ghiglieri:2017gjz} the authors of which use 
a different starting point by making an ansatz with a non-equilibrium density
matrix which contains the chemical potentials from the very start, even
though we differ at intermediate steps.

We have computed the leading order correction
of departures from equilibrium of the charges
to the dispersion relation of the sterile neutrinos in 
the symmetric phase.
There we have considered only the leading order contribution from Standard
Model couplings.
Our equations can be applied
to low-scale leptogenesis and to sterile-neutrino dark matter production.

\bigskip 

\noindent {\bf  \large Acknowledgments } We would like to thank 
Alexander~Klaus and Peter~Reimann for useful discussions,
and Mikko~Laine for valuable comments on the manuscript.
This work was funded in part by
the Deutsche Forschungsgemeinschaft (DFG, German Research
    Foundation) – Project number 315477589 – TRR 211.

 \appendix

\global\long\def\theequation{\thesection.\arabic{equation}}

\section{Perturbative solution of the equations of motion for fluctuations}
\label{a:pert}

The fluctuations of $ y _ a $ satisfy~(\ref{eom}) with an additional 
Gaussian noise $\zeta$, and a $ y $-in\-de\-pen\-dent term that does not
play a role once we consider~\eqref{eom} for departures from equilibrium. 
We solve the equation of motion by one-sided Fourier transformation.
Neglecting non-linear terms and expanding 
as in~(\ref{ypert}), one obtains~(\ref{y0}). 
Now inserting~\eqref{ypert} up to $ y ^ { ( 1 ) } $ into~(\ref{eom}),
including the term with  $ \gamma _ {abc} $ and 
dropping the one with 
$ \gamma_{abcd}  $, we obtain
\begin{align}
  y _ a ^ { + ( 1 ) } ( \omega ) 
   = 
  - \frac 12\left [ ( - i \omega +\gamma )^ { - 1 } \right  ]_ { a b } \;  \gamma _ { b c d } 
    \int 
    \frac { d \omega ' } { 2 \pi } y _ c ^ { + ( 0 ) } ( \omega ' )
    y _ d ^ { + ( 0 ) } ( \omega - \omega ' )
  \label{y1}
  ,
\end{align}
where we have used that
\begin{align}
  y _ a ( t ) = \int 
  \frac{d \omega '}{ 2 \pi } e ^ { - i \omega ' t } y _ a ^ + ( \omega ')
  \label{inverse}
  .
\end{align}
Considering now the correlator $ \mathcal C ^ + _ { a ( b c ) } $ and
inserting~(\ref{y0}) yields averages like $ \langle \zeta y (0) \rangle $ which vanish.
We then consider frequencies much
larger than the rates $ \gamma _ { ab } $, approximating
$ \left [ ( - i \omega +\gamma )^ { - 1 } \right  ]_ { a b } \approx i \delta _ { a b } ( \omega + i
0 ^ + ) ^ { - 1 } $.
The disconnected
contribution vanishes, and we obtain
\begin{align}
  \mathcal C ^ + _ { a ( b c ) } ( \omega )
  \ni
  -
  \frac { i } { 2 \omega } \, \gamma _ { a j k }
  \left[ \Xi _ { j b } \, \Xi _ { k c } + \Xi _ { j c } \, \Xi _ { k b } \right] 
  \int 
  \frac{ d \omega ' } { 2 \pi } 
  \frac { i } { ( \omega ' + i 0 ^ + ) }
  \frac { i } { ( \omega - \omega ' + i 0 ^ + ) }
  \label{cpabc}
\end{align}
along with a contribution from $ \gamma _ { ab } $.
Closing the contour in the upper half-plane gives only a contribution from the second pole and
using the symmetry $ \gamma _ { ajk} = \gamma _{ akj} $,
we arrive at~(\ref{con}).

The perturbation caused by $ \gamma _ { abcd} $ is obtained by
inserting~(\ref{ypert}) into~(\ref{eom}), this time keeping also $ y ^ { ( 2 ) }$.
Having obtained already
$ y ^ { + ( 0 ) } $ and $ y ^ { + ( 1 ) } $ we can
now solve for $ y ^ { + ( 2 ) } $, which reads
\begin{align}
  y _ a ^ { + ( 2 ) } ( \omega ) 
  =&
  -
   \frac{1}{3!}
    \left [ ( - i \omega +\gamma )^ { - 1 } \right  ]_ { a b } \; 
    \gamma _ { b c d e } 
    \nonumber\\
    &
   \times 
    \int 
    \frac { d \omega ' } { 2 \pi } 
    \int
     \frac { d \omega '' } { 2 \pi } 
    y _ c ^ { + ( 0 ) } ( \omega ' ) y _ d ^ { + ( 0 ) } ( \omega '' )
    y _ e ^ { + ( 0 ) } ( \omega - \omega ' - \omega '' )
  \label{y2}
  .
\end{align}
Considering the same limit $ \omega \gg \gamma $, we find contributions of orders
$ 1 $, $ \gamma _ { a b } $, $ \gamma _ { abc } $ and $ \gamma _ {abcd } $ in 
the correlator~(\ref{cabcd}).
The $ \order (1) $ contribution is time-independent and does not contribute to the real part.
The contributions from $ \gamma _ { a b}  $ and 
$ \gamma _ { abc } $ are obtained in the same manner as before.
The contribution from $ \gamma _ { abcd } $ reads
\begin{align}
  \mathcal C ^ + _ { a ( b c d ) } ( \omega )
  \nonumber
  \ni - 
  \frac { i } { 3! \, \omega } \, \gamma _ { a j k l } \big[
    \Xi _ { j b } \, \Xi _ { k c } \, \Xi _ { l d }   
& +   \Xi _ { j b } \, \Xi _ { k d } \, \Xi _ { l c }   
+   \Xi _ { j c } \, \Xi _ { k b } \, \Xi _ { l d }      
\\
+ \; \Xi _ { j c } \, \Xi _ { k d } \, \Xi _ { l b }         
& +   \Xi _ { j d } \, \Xi _ { k b } \, \Xi _ { l c }   
+   \Xi _ { j d } \, \Xi _ { k c } \, \Xi _ { l b }     
  \big]
  \nonumber
  \\
  \times
 \int 
  \frac{ d \omega ' } { 2 \pi }
  &
  \frac{ d \omega '' } { 2 \pi } 
  \frac { i } { ( \omega ' + i 0 ^ + ) } \frac { i } { ( \omega '' + i 0 ^ + ) } \frac { i } { ( \omega - \omega ' - \omega '' + i 0 ^ + ) }
  \label{C4}
\;  .
\end{align}
Because $ \mathcal C _ { a ( b c d ) } $ contains only the connected pieces,
no contractions like $ \Xi _ { j k } \Xi _ { l b } \Xi _ { c d } $ 
appear in the square brackets in~\eqref{C4}, 
and because of~\eqref{Xihier} 
we can neglect
the connected part of the six-point function
$ \langle y _ b y _ c y _ d
          y _ j y _ k y _ l \rangle $.
Making use of the symmetry of $ \gamma _ { ajkl } $ under
permutations of the last three indices,~\eqref{C4} can be solved for the rate
coefficient. Carrying out the integrals and 
collecting the contributions from $ \gamma _ { ab}$ and $ \gamma _ { abc} $ eventually leads to~(\ref{g4}).

\section{Smeared occupancies}
\label{a:smear}

The fluctuations of the occupancies are not small,
$ \Xi _ { f  f } \sim \df $. In order for
the approach in sections~\ref{s:langevin} and~\ref{s:mic}
to be applicable, we consider occupancies
averaged over a certain momentum space region $ \Omega  _ { \vec k } $ 
around $ \vec k $, 
\begin{align}
   F _ { \vec k } 
   \equiv
   \frac { ( 2 \pi  ) ^ 3 } 
      { V | \Omega  _ { \vec k } | }
   \sum _ { \vec p \in 
     \Omega  _ { \vec k } }
   f _ { \vec p } 
   \label{smear}
   .
\end{align} 
The volume of this region $ | \Omega  _ { \vec k } | $ 
is taken to be independent of the spatial volume $ V $, with
$ (2\pi)^3 / V \ll | \Omega  _ { \vec k } | \ll T^3 $.
The susceptibilities~\eqref{Xihigh}
of $ F _ { \vec k } $ are of order
$
( V | \Omega  _ { \vec k } | ) ^{ - n + 1 }
$.
Now we have $  \sqrt{ \Xi _ { {\scriptscriptstyle F} _ { \vec k } 
                                      {\scriptscriptstyle F} _ { \vec k } } }
\ll \dF _ { \vec k } $, and the assumption~\eqref{Xihier}
is satisfied.

We should also consider smeared occupancies in the microscopic
correlators appearing in section~\ref{s:hexp}. 
However, since the volume $ | \Omega _ { \vec k } | $ is small 
compared to characteristic momentum scales over which the correlators
in~\eqref{kubo}--\eqref{gamma3} vary, they can to a good approximation be replaced
by
\begin{align}
\tpf _ { F _ { \vec k } ( F _ { \vec k '} X \cdots X ) } ^ { \rm ret } ( \omega )
=
\delta _{ \vec k \vec k ' }
\frac { (2\pi) ^3 } { V | \Omega _ { \vec k } | }
\tpf _ { f _ { \vec k } ( f _ { \vec k } X \cdots X ) } ^ { \rm ret } ( \omega )
\label{FFff}
,
\end{align}
so that the dependence on $ V | \Omega _ { \vec k } | $ drops out when
plugging~\eqref{FFff} and $ \Xi _ { F F } $ into the master formula~\eqref{kubo},
and one can effectively use the unsmeared occupancies $ f $.

\section{Green's functions at finite temperature and che\-mical potentials}
\label{a:ana}

Here  we present slight generalizations of  some relations
in~\cite{Laine:2016hma} for imaginary 
time correlators which we use in our calculation. 
In the presence of one conserved charge $ Q $ 
the 2-point function of operators $ A $, $ B $ reads 
\begin{align}
  \tpf    _ { AB } ( -i \tau   ) 
   \equiv Z ^{ -1 } {\rm tr} \!\left \{  e ^{ 
      \beta  ( \mu  Q - H ) }
    \T \left [  A ( -i \tau   ) B ( 0 ) \right ] \right \} 
  \label{imaginary} 
\end{align} 
with $ \beta \equiv 1 / T $, the partition function $ Z \equiv 
{\rm tr}  \, e ^{ 
      \beta  ( \mu  Q - H ) } $, and the  time ordering 
  $ \T $ 
is with respect to $ \tau  $.
\eqref{imaginary} is defined for
$ -\beta  \le \tau  \le \beta  $. 
We assume that $ A $ carries a definite charge, 
\begin{align}
  [ Q, A ] = q  _ A  A 
  \label{QA}
  .
\end{align}
Then 
\begin{align}
  \tpf _ { AB } ( t + i \beta  ) = \pm e ^{ -\beta  \mu  _ A }
  \tpf _ { AB } ( t )
  \label{period} 
\end{align} 
where
the chemical potential of $ A $ is defined as
\begin{align}
  \mu  _ A \equiv q _ A  \mu 
  \label{muA}
  .
\end{align} 
Therefore the function $ e ^{ - \mu  _ A \tau } \tpf _ { AB } ( -i \tau ) $
is (anti-) periodic, and can be expanded in a Fourier series with
coefficients
\begin{align}
  \tpf _ { AB } ^{ \rm M } ( i \omega  _ n ) 
  \equiv
  \int  _ 0 ^ \beta  \! d \tau  \; e ^{ ( i \omega _ n -  \mu  _ A ) \tau  } 
  \tpf _ { AB } ( -i \tau  )
  \label{matsubara} 
  .
\end{align}
(\ref{matsubara}) can be analytically continued to arbitrary complex
frequencies off the real axis, and we denote the resulting function by
$ \tpf _ { AB } ^{ \rm M } $.  We need to calculate the retarded
correlator
\begin{align}
  \tpf ^ {\rm ret } _ { AB } ( \omega  )
  = 
  i   \int _ 0 ^ {\infty  }\! d t \;
  e ^{ i \omega  t } 
  \langle [ A ( t ) , B ( 0 ) ] _ \mp \rangle 
  \label{retarded} 
\end{align} 
where $ \omega $ is real.
It can be analytically continued to the complex plane. We denote
the resulting function by $ \tpf  _ { AB } $, and then we have
$ \tpf ^ {\rm ret } _ { AB } ( \omega  ) = \tpf _ { AB } ( \omega  + i 0 ^ + ) $.
The two analytic continuations  are related by
\begin{align}
   \tpf _ { AB } ^{ \rm M } ( \omega  ) 
   =
   \tpf _ { AB }( \omega  - \mu  _ A ) 
   \label{matsret} 
   .
\end{align}

\section{Cancellation of the rates
         \texorpdfstring{$ \gamma _{\scriptscriptstyle{fff}}$}{gamma fff} and
         \texorpdfstring{$ \gamma _{\scriptscriptstyle{ffff}}$}{gamma ffff}
         }
\label{a:gfff}

Here we demonstrate that the coefficients $ \gamma _ { f f f } $
and $ \gamma _ { f f f f } $
in the equation of motion for $ f $ vanish at order $ h ^ 2 $.
For simplicity we will assume that the
$ a $ and $ a ^ \dagger $ appearing in all occupancies in this appendix
correspond to sterile neutrino generations satisfying~\eqref{degen}.
First consider $ \gamma _ { f f  f} $.
According to~\eqref{gamma2} it consists of two pieces
containing only $ f $ operators (suppressing momentum indices~$ \vec k $),
\begin{align}
\gamma _ { f ^ a f ^ b f ^ c } 
=
T \omega  
  \, {\rm Im} \left [ \tpf ^ {\rm     ret}_{f ^ a ( f ^ d  f ^ e ) }(\omega)
    - \tpf ^ {\rm     ret}_{f ^a f^g }(\omega) (  \Xi^{-1})_{f^g f^f}
    \,
    \Xi  _ { f^f f^d f^e }
    \right ]
    (\Xi^{-1})_{f^d f^b} (\Xi^{-1})_{f^e f^c}
\label{gfff}
.
\end{align}
Classically, the kinematic
variables commute at equal times, which is not the case in the microscopic
theory. Therefore one should replace the product $ f ^ d f ^ e $ in~\eqref{gfff}
by its symmetrization $ \{ f ^ d , f ^ e \} / 2 $.
In turn we demonstrate the cancellation of the terms with the ordering as
in~\eqref{gfff},
the one with $ d \leftrightarrow e $ is analogous.

We obtain the generalized susceptibility
\begin{align}
\Xi _ { f ^ a   f ^ b   f ^ c }
=
f _ { \rm F }
  \left[ 1 - f _ { \rm F } 
   \right]
\Big\{ \left[ 1 - f _ { \rm F } 
  \right] 
        \, {\rm tr }  \! \left( T ^ a T ^ b T ^ c \right)
        - f _ { \rm F } 
        \; {\rm tr }  \! \left( T ^ a T ^ c T ^ b \right)
   \label{chifff} 
\Big\}
,
\end{align}
where $ f _ { \rm F } = f _ { \rm F }  ( E _ \vec k ) $.
The mass in $ E _ { \vec k } $ is 
one of the relevant (nearly) degenerate masses, and a change in this mass gives 
only a correction of order $ h ^2 \dM^2$, which we neglect.
We now consider the correlators as a function of imaginary time $t=-i\tau$, before Fourier
transformation and analytic continuation to real frequency. Then, 
using~\eqref{Tnorm}
and the susceptibility~\eqref{chiff}, the second expression contains a term
\begin{align}
\tpf _{f ^a f^g }( t ) (  \Xi^{-1})_{f^g f^f}
    \Xi  _ { f^f f^d f^e }
    \ni
    \frac {1}{4 V E _  {\vec k } }  T ^ a _ { i j }
    &
    \left\{ \left[ 1 - f _ { \rm F } ( E _ { \vec k } ) \right] 
        \left( T ^ d T ^ e \right) _ { l m }
        - f _ { \rm F } ( E _ { \vec k } )
        \left( T ^ e T ^ d \right) _ { l m }
    \right\}
    \nonumber    
    \\
\int _ 0 ^ { 1/T} \! d \tau _ 1 \, d \tau _ 2
\Big\{
&
\big\langle a _ q  ^ \dagger ( t _ 1 ) a _ m ( 0 ) \big\rangle
\big\langle \T a _ r ( t _ 2 ) a _ i ^ \dagger ( t ) \big\rangle
\big\langle a _ j  ( t ) a _ l ^ \dagger ( 0 ) \big\rangle
\nonumber
\\
-
&
\big\langle \T a _ q  ^ \dagger ( t _ 1 ) a _ j ( t ) \big\rangle
\big\langle a _ r ( t _ 2 ) a _ l ^ \dagger ( 0 ) \big\rangle
\big\langle a _ i ^ \dagger ( t ) a _ m ( 0 ) \big\rangle
\Big\}
\nonumber\\
\int \! d ^ 3 x_1 \, d ^ 3 x _ 2 \;
&
\overline u _ { q +}  h _ { q \alpha }
\big\langle J _ \alpha ( t _ 1, x_1 ) \bar J _ \beta ( t _ 2 ,x_2) \big\rangle
h ^ \dagger _ { \beta r } u _ {r +}
\label{fff2}
.
\end{align}
To meaningfully define the object $\tpf _{f ^ a ( f ^ d  f ^ e ) }(t )$ in the
path integral over the sterile neutrino fields, we separate the quantities
at  $ t = 0 $
by replacing $ f ^ d ( 0 ) $ by $ \lim \limits _ { t' \to 0 ^ + } f ^ d ( t' ) $,
taking the limit in the end when the ambiguities have resolved, which
is the case after the expectation value has been reduced 
using Wick's theorem. The
equivalent expression to the one in~\eqref{fff2} now contains 8 terms which
fall in either of the following two categories: (i)~A~product of 4 two-point
functions of operators $ a $ and $ a ^ \dagger $, in which exactly one operator
is at time $ t = 0 $, and the other one at a time that is integrated over, or
(ii)~a product of 3 two-point functions in which the operators are at different times,
multiplied by one $ f _ { \rm F }  $ or $ \left[ 1 - f _ { \rm F } \right]$.%
\footnote{Here the time-ordering decides which one of the two expressions
  $ f _ { \rm F} $ or $ \left[ 1 - f _ { \rm F } \right]$ is
  generated, and the temporal separation of $ f ^ d $ and $ f ^ e $ plays a role.}
The 4 expressions of type~(i) are $t$-independent, since operators at $ t $ are
always to the left of those at time $ 0 $ so that they are
not affected by time ordering,
and the time evolutions of $ a _ j ( t ) $ and $ a ^ \dagger _ i ( t ) $
cancel up to effects of order $ h ^ 2 \dM^2 $, which we neglect.
Time-independent parts do not contribute to our master formula.
The remaining four terms of type (ii) are canceled by the terms in~\eqref{fff2}.
The other contributions which we have not written in~\eqref{fff2}
are canceled in the same way, and we obtain $ \gamma_{fff}=0$.

The master formula for the coefficient $ \gamma _ { f f f f } $ also
contains only $ f $ operators at order~$ h ^ 2 $. The contribution from
$ \gamma _ { f f f } $ vanishes, and the two remaining terms read
\begin{align}
  \gamma  _ { f ^ a  f ^ b  f ^ c  f ^ d } 
  = \nonumber
  T \omega  
  \, {\rm Im} \Big [  \tpf ^ {\rm     ret}_{ f ^ a ( f ^ e f^ f f ^g ) }(\omega)
   & - 
     \tpf ^ {\rm     ret}_{f ^ a f ^ h }(\omega)
    (  \Xi^{-1})_{ f ^ h f ^ i } \,
     \Xi  _ { f ^i f ^ e f ^ f f ^ g } 
        \Big ]
        \nonumber
    \\
   & \quad\quad\quad\quad\quad
   \times    
   (\Xi^{-1})_{f ^ e  f ^ b} (\Xi^{-1})_{f ^ f   f ^ c} (\Xi^{-1})_{f ^ g   f ^d}
        \label{gffff}
  .
\end{align}
Using~\eqref{Tnorm}, we obtain the generalized susceptibilities
\begin{alignat}{2}
\Xi _ { f ^ a f ^ b f ^ c f ^ d }
=
f_{\rm F}\left[1-f_{\rm F}\right]
\Big\{
&\left[ 1 - f _ { \rm F } \right] ^2
        \,&& {\rm tr }  \! \left( T ^ a T ^ b T ^ c T ^ d \right)
- f _ { \rm F } \left[ 1 - f _ { \rm F } \right]
        \, {\rm tr }  \! \left( T ^ a T ^ b T ^ d T ^ c \right)
\nonumber
\\
- f _ { \rm F } &  \left[ 1 - f _ { \rm F } \right]
        \,&& {\rm tr }  \! \left( T ^ a T ^ c T ^ b T ^ d \right)
- f _ { \rm F } \left[ 1 - f _ { \rm F } \right]
        \, {\rm tr }  \! \left( T ^ a T ^ c T ^ d T ^ b \right)
\nonumber
\\
- f _ { \rm F } & \left[ 1 - f _ { \rm F } \right]
        \,&& {\rm tr }  \! \left( T ^ a T ^ d T ^ b T ^ c \right)
+ f _ { \rm F } ^2 
        \;\, {\rm tr }  \! \left( T ^ a T ^ d T ^ c T ^ b \right)
        \Big\}
\label{chiffff}
.
\end{alignat}
The cancellation is now analogous to the one above,
after replacing $ f ^ e ( 0 ) f ^ f ( 0 ) $ by \linebreak
$ \lim \limits _ { t'' \to 0 ^ + } \lim \limits _ { t' \to 0 ^ + }
    f ^ e ( t'' )  f ^ f ( t' ) $ with $ t'' > t'$.

\bibliographystyle{jhep}
\bibliography{references}

\end{document}